\documentclass[aps,pre,floatfix, twocolumn,preprintnumbers,superscriptaddress,amsmath,amssymb,10pt,a4paper]{revtex4-2}

\usepackage{geometry}
\usepackage{multirow}
\usepackage{blkarray}
\usepackage{mathtools}
\usepackage{graphicx}
\usepackage{dcolumn}
\usepackage{bm}
\usepackage{makecell}
\usepackage{xcolor}
\usepackage{enumerate}
\usepackage[caption=false]{subfig}
\usepackage{float}
\usepackage[unicode=true,pdfusetitle,
 bookmarks=true,bookmarksnumbered=false,bookmarksopen=false,
 breaklinks=false,pdfborder={0 0 1},backref=false,colorlinks=true, hypertexnames=false]
 {hyperref}
\hypersetup{linkcolor=blue,urlcolor=blue,citecolor=blue}
\usepackage{natbib}
\usepackage{physics}
\usepackage{placeins}
\usepackage{titlesec}

\newcommand{\be}{\begin{equation}}
\newcommand{\ee}{\end{equation}}
\newcommand{\bd}{\begin{displaymath}}
\newcommand{\ed}{\end{displaymath}}
\newcommand{\BE}{\begin{eqnarray}}
\newcommand{\EE}{\end{eqnarray}}

\newcommand{\bx}{\ensuremath{\mathbf{x}}}

\newcommand{\avg}[1]{\left\langle{#1}\right\rangle}

\newcommand{\icol}[1]{
  \left(\begin{smallmatrix}#1\end{smallmatrix}\right)%
}
\newcommand{\Tanh}[1]{\tanh\left({#1}\right)}

\newcommand{\nocontentsline}[3]{}
\let\oldaddcontentsline=\addcontentsline
\let\addcontentsline=\nocontentsline

\begin{document} 

\title{\texorpdfstring{$q$}{}-deformed evolutionary dynamics in simple matrix games}

\author{Christopher R. Kitching}
\email{christopher.kitching@manchester.ac.uk}
\affiliation{Department of Physics and Astronomy, School of Natural Sciences, The University of Manchester, Manchester M13 9PL, UK}
\author{Tobias Galla}
\email{tobias.galla@ifisc.uib-csic.es}
\affiliation{Instituto de F\'isica Interdisciplinar y Sistemas Complejos, IFISC (CSIC-UIB), Campus Universitat Illes Balears, E-07122 Palma de Mallorca, Spain}

\date{\today}

\begin{abstract}
We consider evolutionary games in which the agent selected for update compares their payoff to $q$ neighbours, rather than a single neighbour as in standard evolutionary game theory. Through studying fixed point stability and fixation times for $2\times 2$ games with all-to-all interactions, we find that the flow changes significantly as a function of $q$. Further, we investigate the effects of changing the underlying topology from an all-to-all interacting system to an uncorrelated graph via the pair approximation. We also develop the framework for studying games with more than two strategies, such as the rock-paper-scissors game where we show that changing $q$ leads to the emergence of new types of flow.

\end{abstract}
\maketitle

\section{Introduction}
Modern game theory was pioneered by mathematician John von Neumann in the 1920s to mathematically model strategic interactions among agents whose decisions are rational \cite{v1928theorie,von1959theory,von2007theory}. In the 1970s John Maynard Smith and George Price built on this idea by introducing evolutionary game theory \cite{smith1973logic, smith1982evolution}. Here, agents do not act rationally, instead each individual carries a particular strategy that is passed on from parent to offspring. Reproduction occurs in proportion to fitness. This was initially conceived as a model of Darwinian evolution in biology, but has also found applications in the social sciences and in economics \cite{Gintis2000}. Originally, evolutionary game dynamics in populations were described mostly by deterministic differential equations, such as the commonly used replicator equations \cite{hofbauer2003evolutionary}. These equations are formally valid for infinite populations.

However, it is now well established that stochastic effects can considerably affect the outcome of evolution (see e.g. \cite{taylor,traulsen2005coevolutionary,traulsen2009stochastic}). There is thus a continuously growing body of work on evolutionary game theory in finite populations. One particular focus is also on networked populations, that is, population in which any one individual can only interact and compete with its immediate neighbours. This has sometimes been referred to as `evolutionary graph theory' \cite{liebermann}. This should should not be misunderstood as a theory of evolving graphs. Despite the terminology, the underlying graph does often not change in time.

The starting point for studies of evolutionary game theory in finite populations and networks is usually an individual-based model. This is a set of rules by which the agents interact and the population evolves. A number of different interaction models have been proposed, see e.g. \cite{Traulsen_et_al_2006,claussen}; a summary can also be found in \cite{bladon}. For analytical convenience the total size of the population is often kept fixed. Here, we focus on the so-called pairwise comparison process mediated via `Fermi functions' \cite{traulsen2006stochastic}. Detailed definitions will be given in Sec.~\ref{sec: model definitions}.

The so-called voter model is a related, but different model of interacting individuals. It was originally introduced by  Holley and Liggett in 1975 to model interacting particle systems \cite{holley1975ergodic}. In the most basic voter model, the population consists of voters which are of binary types (`opinions'), 0 or 1, connected in some way through a network. In each step of the dynamics one randomly chosen individual adopts the opinion of a randomly chosen neighbour. Many extensions and variations of this model have been proposed and studied \cite{granovsky1995noisy, mobilia2007role, masuda2013voter, castellano2009statistical,redner}. One such variation is the so-called `$q$-voter model', in which $q$ neighbours need to agree with one another in order for a voter to change opinion \cite{castellano2009nonlinear,mobilia,ramirez}.

In this paper we seek to combine the ideas of evolutionary games and the $q$-voter model into `$q$-deformed' evolutionary game dynamics. We consider a population of individuals who each carry a particular strategy, and the population then evolves following rules similar to those in conventional models of evolutionary game theory, in particular the probability for an agent to change state depends on payoff. However, before a change can occur, an agent must consult with $q$ of its neighbours. As in the $q$-voter model a change of that agent can then only occur if all those $q$ neighbours are of the same type. As we will show this can significantly affect the evolutionary flow in strategy space, and quantities such as fixation probabilities and times in finite populations.

The remainder of the paper is organised as follows. In Sec.~\ref{sec: model definitions} we introduce the model of $q$-deformed evolutionary game dynamics for $2\times 2$ games (two-strategy two-player games). In Sec.~\ref{sec: rate equations for infinite populations} we investigate the rate equations for the $q$-deformed dynamics in infinite all-to-all populations. We study the fixed points and their stability, and show that the type of evolutionary flow changes as a function of $q$. In Sec.~\ref{sec: fixation times} we look at fixation times and probabilities in finite all-to-all systems. In Sec.~\ref{sec: graphs} we study the effects of changing the underlying topology from a complete graph to an uncorrelated graph. Analytical results for $q$-deformed dynamics are obtained within the pair approximation. In Sec.~\ref{sec: multi strategy} we study games with more than two strategies. We investigate cyclic games (motivated by the familiar rock-paper-scissors game) as an example and show that $q$-deformation can induce new behaviour, in particular the emergence of stable limit cycles.  Finally in Sec.~\ref{sec: games without replacement} we look at a modified version of the model, where we select the $q$-neighbours without replacement. We summarise our results in Sec.~\ref{sec:conclusions}, and give an outlook on possible future work.

\section{Model definitions} \label{sec: model definitions}
\subsection{Payoff matrix}
We consider a population of $N$ agents, who are each of type $A$ or $B$. For the time being we always consider a population with all-to-all interaction.  

The game is defined by the $2\times 2$ payoff matrix
\begin{equation}
    \begin{blockarray}{cccc}
     & & A & B  \\
    \begin{block}{cc(cc)}
        A~ & & a & b \\
        B~ & & c & d \\
    \end{block}
    \end{blockarray} \hspace{3mm} . \label{eq main: payoff matrix}
\end{equation}
An $A$ agent interacting with another $A$ agent thus receives payoff $a$, and it receives $b$ if interacting with an agent of type $B$. Likewise a $B$ agent receives $c$ when interacting with an $A$, and $d$ when interacting with a $B$. 

In finite systems we characterise the state by the number of type $A$ agents $i$, or, equivalently, by the fraction of $A$ agents in the population, $x=i/N$. The expected payoff for type $A$ and $B$ agents respectively is then
\begin{gather} 
\begin{aligned} 
    \pi_{A}(x) &= ax+b(1-x), \\
    \pi_{B}(x) &= cx + d(1-x). \label{eq main: avg payoff}
\end{aligned}
\end{gather}
We can define the payoff difference as
\begin{equation}
    \Delta\pi(x) = \pi_{A}(x) - \pi_{B}(x) = ux+v, \label{eq main: payoff diff}
\end{equation}
where,
\begin{gather} \label{eq: uv}
\begin{aligned} 
    u &= (a+d)-(b+c), \\
    v &= b-d. 
\end{aligned}
\end{gather}
For evolutionary processes relying only on payoff differences it is thus possible to represent the payoff matrix in Eq.~(\ref{eq main: payoff matrix}) with only two variables: $u$ and $v$. 

\subsection{$q$-deformed evolutionary dynamics}
At each time step we randomly choose an agent for update, say it is of type $A$. We then choose $q$ agents from the population (with replacement). If all those agents are of type $B$, then the $A$ switches to $B$, with a probability $g^-$ that depends on the payoffs of type $A$ and $B$ agents. Similarly, we write $g^+$ for the probability for a $B$ to switch to $A$ if $q$ other randomly selected agents also are of type $A$.

There are many possible choices for the functions $g^\pm$. One such choice is to make $g^\pm$ proportional to the payoff of the type that is being switched to, i.e.
\begin{equation}
    g^{\pm}(x) = \frac{\pi_{A/B}(x)}{\text{max}\{a,b,c,d\}}. \label{eq: g const}
\end{equation}
This choice is is only valid if all entries in the payoff matrix are positive. The denominator ensures that $g^\pm(x)\leq 1$. In this way, if type $B$ agents have a large average payoff, then type $A$ agents are likely to switch during interactions with them. But it is still possible for an agent to switch to a type with a lower expected payoff than its current type. A second popular choice involves non-linear Fermi functions $g^\pm$ \cite{blume1993statistical, traulsen2006stochastic}, 
\begin{equation}
    g^{\pm}(x) = \frac{1}{1+e^{\mp\beta\Delta\pi(x)}}. \label{eq main: g fermi}
\end{equation}
These depend on the payoff difference. The parameter $\beta\geq 0$, known as the intensity of selection, reflects the uncertainty in decision-making processes. In the strong selection limit, $\beta\rightarrow\infty$ an individual will always switch to the type with the higher expected payoff. For any finite $\beta$ the reverse process occurs with non-zero probability.  For $\beta = 0$ one has $g^\pm =\frac{1}{2}$, i.e. neutral selection. Payoffs then have no effect. Since the parameter $\beta$ only appears in combination with $u$ and $v$, we will absorb it into those parameters, effectively setting $\beta = 1$.

The rates at which the type $A$ and $B$ agents switch are given by,
\begin{gather} \label{eq: Tplusminus}
\begin{aligned} 
    T^{+}(x) &\equiv T_{B\rightarrow A}(x) = N(1-x)x^{q}g^{+}(x), \\
    T^{-}(x) &\equiv T_{A\rightarrow B}(x) = Nx(1-x)^{q}g^{-}(x),
\end{aligned}
\end{gather}
respectively.
The first expression can be understood as follows. The agent chosen for reproduction is of type $B$ with probability $1-x$. The probability that $q$ randomly chosen neighbours (with replacement) in this all-to-all population are all of type $A$ is $x^{q}$. The type $B$ agent switches to type $A$ with probability $g^+(x)$. We will sometimes write $T_i^\pm$ instead of $T^\pm(x)$, noting that $x=i/N$.

We measure time in units of generations, the rates are therefore proportional to the size $N$ of the population. While the above motivation of the model implicitly assumes that $q$ is a positive integer, the continuous-time rates in Eq.~(\ref{eq: Tplusminus}) are mathematically meaningful for all $q>0$. We thus treat $q>0$ as a real-valued model parameter, similar to what is done in the literature on the $q$-voter model \cite{timpanaro2014exit,mobilia,mellor2016characterization,jed_qvm,sznajd_qvm,ramirez}. For $q=1$ the setup reduces to that of conventional evolutionary game theory in finite populations (see e.g. \cite{traulsen2009stochastic,traulsen2005coevolutionary,altrock2009fixation,bladon}). We will refer to choices $q\neq 1$ as a `$q$-deformation' of the dynamics.

It is easy to understand that $q$-deformation favours strategies that are in the minority when $q<1$, or in majority respectively, for $q>1$. To see this it is useful to focus on the case with no selection ($g^\pm\equiv 1/2$ in our notation). If the frequency of strategy $A$ is $x$, the rate of change from $A$ to $B$ is proportional to $x(1-x)^q$ and that for the reverse change proportional to $x^q (1-x)$. For $q=1$ these terms balance for all $x$. This is the case of the conventional voter model, where there is no deterministic drift. For $q<1$ we have $x(1-x)^q < (1-x) x^q$ when $x<1/2$. Thus for $q<1$ strategy $A$ is favoured when it is in the minority, and disfavoured when more than half of population is of type $A$. The direction of the flow reverses for $q>1$.
\section{Rate equations for infinite populations} \label{sec: rate equations for infinite populations}

\subsection{$q$-deformed rate equations}
For large populations the average dynamics of the population is approximated by a $q$-deformed version of the conventional rate equations in an infinite population. This is explained further in the  Supplemental Material (SM) \cite{supp}, see in particular Sec.~\ref{appendix: master eq derivation}. We have
\begin{align}
    \dot{x} 
    &= (1-x)x^{q}g^{+}(x)  -x(1-x)^{q}g^{-}(x). \label{eq: q-deformed replicator}
\end{align}
 The quantity $x$ in this equation is really $\avg{x}$, an average over individual realisations of the dynamics. However to ease notation we will write $x$ throughout. Also note if we choose $g^\pm$ as in Eq.~(\ref{eq: g const}), then after absorbing the constant by re-scaling time we find,
\begin{equation}
    \dot{x} = (1-x)x^{q}\pi_{A}(x)-x(1-x)^{q}\pi_{B}(x).
\end{equation}
This is a $q$-deformed variant of the conventional replicator equations.
Setting $q=1$ reduces this to the standard replicator equation for $2$-player $2$-strategy games \cite{nowak2006evolutionary},
\begin{align}
    \dot{x} &= x(1-x)\left[\pi_{A}(x)-\pi_{B}(x)\right].
\end{align}

Given an initial distribution of type $A$ and $B$ agents in a population, Eq.~(\ref{eq: q-deformed replicator}) determines how the composition of the population changes with time. The system will eventually reach a fixed point. The location of the fixed point depends on $u$ and $v$, and (potentially) on the initial condition.  Our first step is therefore an analysis of the fixed points of the $q$-deformed rate equations and of the stability of these fixed points. As a benchmark we first summarise the well-known results for the case $q=1$.

\subsection{The case \texorpdfstring{$q = 1$}{}}
The simplest case is that of $q=1$, where Eq.~(\ref{eq: q-deformed replicator}) reduces to
\begin{equation}
    \dot{x} = x(1-x)[g^{+}(x) - g^{-}(x)].  \label{eq main: q=1 replicator}
\end{equation}
In this case, there are always two fixed points at the boundaries, $x^{*}=0,1$. There is also potentially a third, interior, fixed point given by the solution of $g^{+}(x^{*})=g^{-}(x^{*})$. For the functions in Eqs. (\ref{eq: g const}) and (\ref{eq main: g fermi}), this equates to solving $\Delta\pi(x^{*}) = 0$, so 
\begin{equation}
    x^{*}=\frac{-v}{u} = \frac{b-d}{c-a+b-d}. \label{eq main: q=1 interior fixed point}
\end{equation}
For some  values of $u$ and $v$ this solution is unphysical ($x^*<0$ or $x^*>1$). We then only have the fixed points at $x^*=0$ and $x^*=1$. In the special case $a=c$ and $b=d$ (neutral selection) we have $\dot{x}=0$, in which case every point $x$ is a fixed point. 

\begin{figure}[htbp]
    \centering
    \captionsetup{justification=justified}
    \includegraphics[scale=0.52]{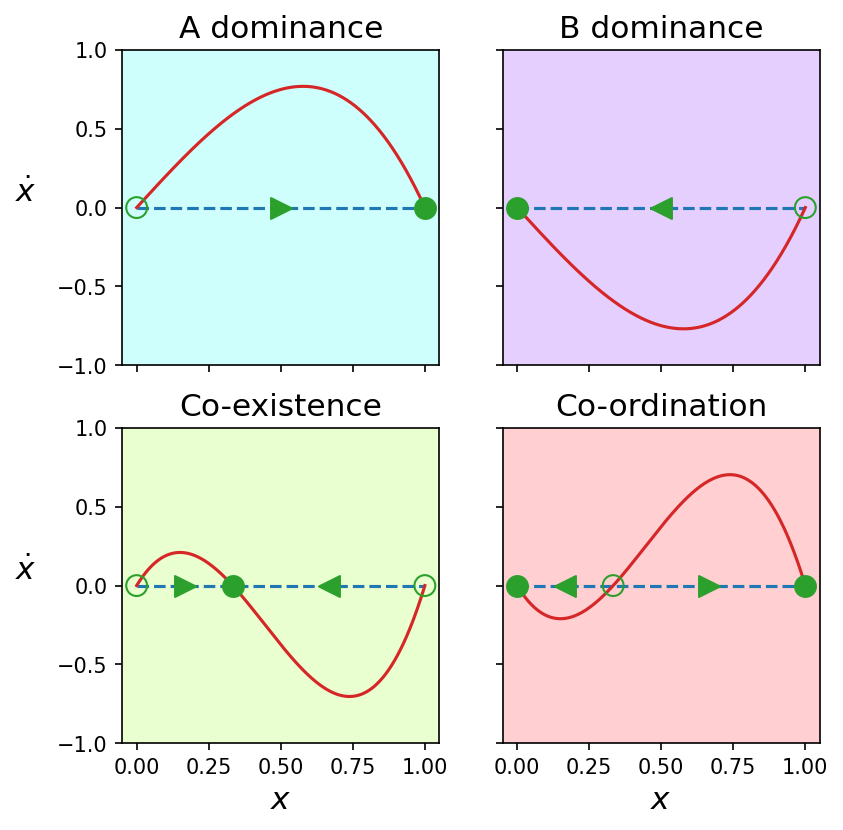}
    \caption{Classifications of the deterministic dynamics for $q=1$, Eq.~(\ref{eq main: q=1 replicator}). Each classification has an associated colour for easy identification in other plots. Arrows indicate the direction of the flow. Filled/empty green circles are stable/unstable fixed points.}
    \label{fig: q=1 games}
    \end{figure}

Given values for $u$ and $v$ we can characterise the type of flow based on the number of fixed points and their linear stability. From Eq.~(\ref{eq main: q=1 replicator}), if $g^{+}(0)>g^{-}(0)$ then $x^{*}=0$ is unstable. If $g^{+}(1)>g^{-}(1)$ then $x^{*}=1$ is a stable fixed point. If an interior fixed point $x^*$ exists, it is stable if
\be
    \frac{\dd}{\dd x}\left[g^{+}(x)-g^{-}(x)\right]\Bigr|_{x^{*}} < 0. \label{eq main: q=1 stable condition}
\ee
These scenarios are summarised in Fig.~\ref{fig: q=1 games}. They are conventionally referred to as the dominance of $A$ or $B$, `co-existence' or `co-ordination', respectively \cite{traulsen2009stochastic}.

\begin{figure}[htbp]
    \centering
    \captionsetup{justification=justified}
    \includegraphics[scale=0.57]{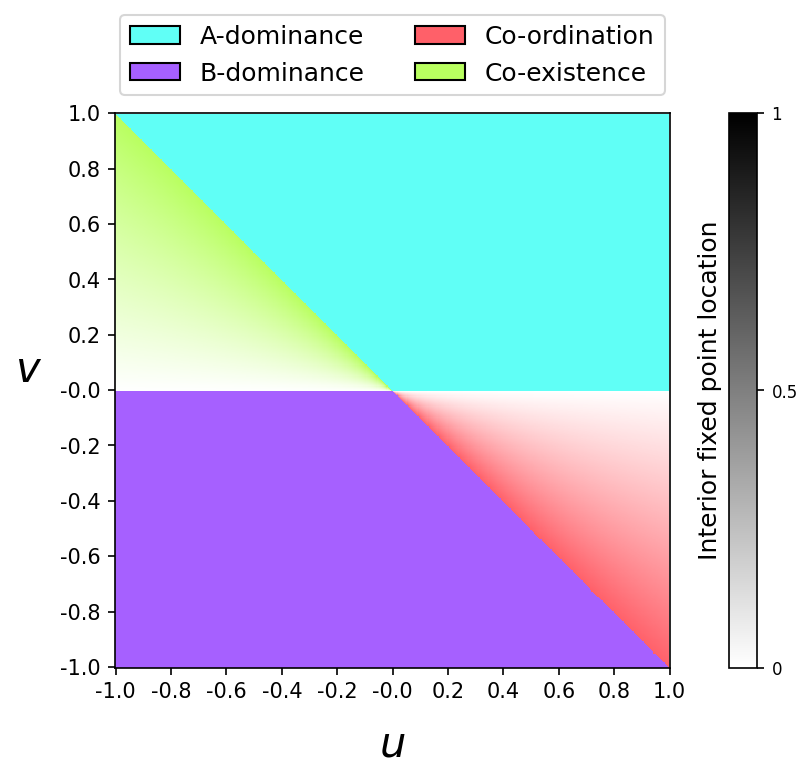}
    \caption{Phase diagram in ($u$,$v$)-space showing the classification regions for evolutionary flow for $q=1$, Eq.~(\ref{eq main: q=1 replicator}). The four colours represent the four possible classifications. The co-existence and co-ordination regions have varying opacity to represent the position of the interior fixed point on the interval $(0,1)$.}
    \label{fig: q=1 phase diagram}
    \end{figure}

We can create a phase diagram in the ($u$,$v$)-space to demonstrate how the evolutionary flow changes when we alter the payoff matrix. Fig.~\ref{fig: q=1 phase diagram} shows how this space divides into four regions with the four types of flow. The diagram is valid for the choices of $g^\pm$ in Eqs.~(\ref{eq: g const}) and (\ref{eq main: g fermi}). In the case of co-existence or co-ordination, the interior fixed point can vary on the interval $(0,1)$ which is represented by the opacity of those regions. The shape of these regions can be explained as follows. We have one interior fixed point when $0<x^{*}<1$ and from Eq.~(\ref{eq main: q=1 interior fixed point}) this implies $0<\frac{-v}{u}<1$. When $u>0$ we find that $v<0$ and $v>-u$, this defines the red region in Fig.~\ref{fig: q=1 phase diagram}. We then evaluate Eq.~(\ref{eq main: q=1 stable condition}) at $x^{*}=\frac{-v}{u}$, to find that $u>0$ means that the interior fixed point is unstable, hence one has co-ordination type flow [Fig.~\ref{fig: q=1 games}]. Similar analyses can be carried out for the other regions. We note here how the co-existence and co-ordination games blend into dominance games as we change $u$ and $v$, since their interior fixed points move closer the the boundaries, eventually being absorbed.

\subsection{The case \texorpdfstring{$q \neq 1$}{}}
We now focus on the case $q\neq 1$ in Eq.~(\ref{eq: q-deformed replicator}). Much like the $q=1$ case we always have fixed points at the boundaries, $x^*=0$ and $x^*=1$. The number of interior fixed points depends on the functions $g^\pm$.

\begin{figure}[htbp]
    \centering
    \captionsetup{justification=justified}
    \includegraphics[scale=0.13]{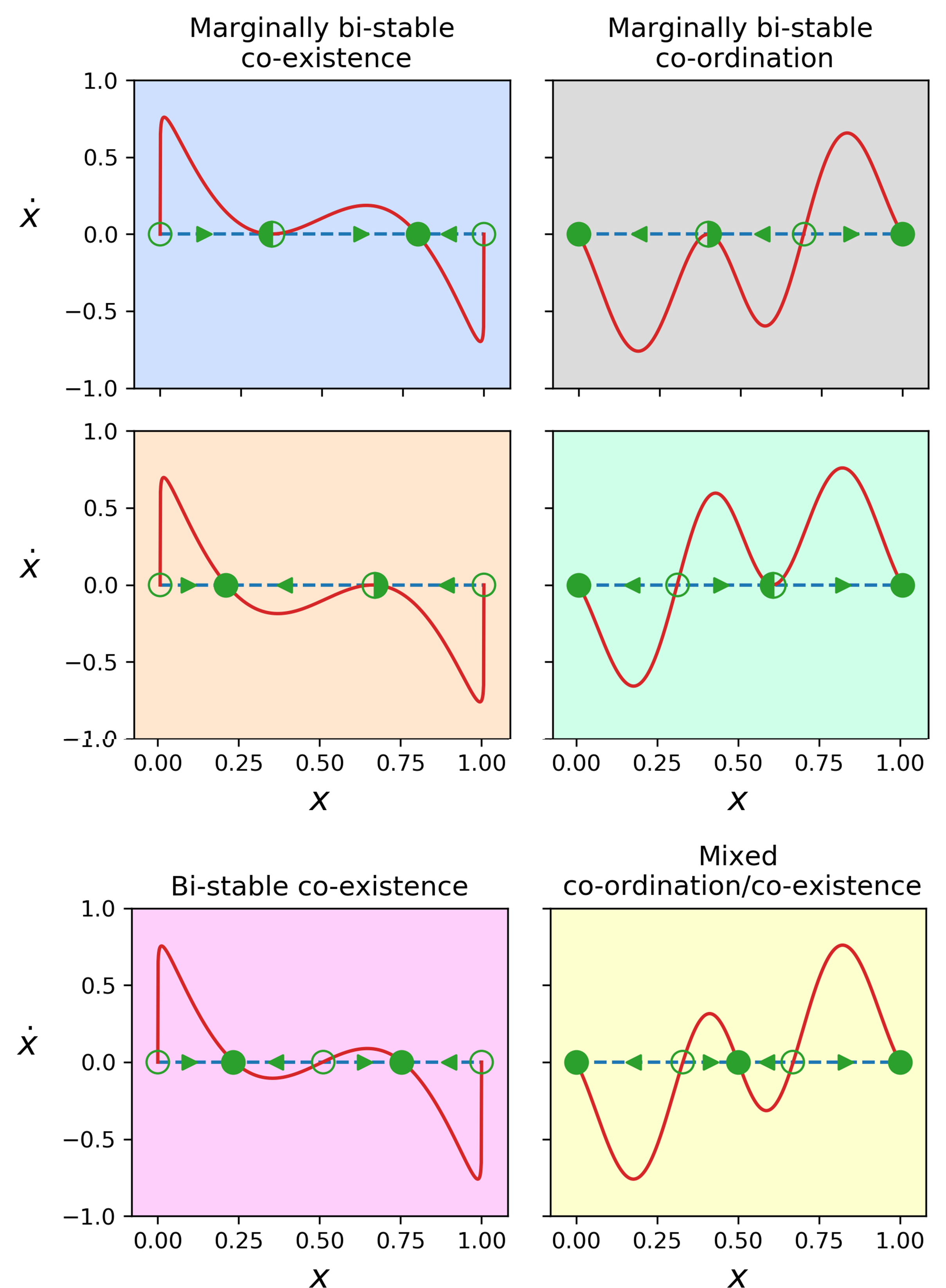}
    \caption{Possible classifications of the $q\neq1$ game dynamical flow in Eq.~(\ref{eq: q-deformed replicator}), with $g^\pm$ Fermi functions as in Eq.~(\ref{eq main: g fermi}). The left column shows $q<1$ co-existence type games, and the right column shows $q>1$ co-ordination type games. Each classification has an associated colour for easy identification in other plots.}
    \label{fig: q/=1 games}
    \end{figure}

All results in this section are for the Fermi function Eq.~(\ref{eq main: g fermi}). We find that there can be between one and three interior fixed points. For $q>1$, we only find co-ordination type flows (i.e. flows with stable fixed points at the boundaries), and for $q<1$ we only find co-existence type flows (i.e. flows with unstable fixed points at the boundaries) [see SM, Sec.~\ref{appendix: number of interior fixed points}]. This is because the $q$-deformation favours majority strategies for $q>1$, while the minority strategy has an advantage for $q<1$. The possible types of flow of Eq.~(\ref{eq: q-deformed replicator}) include standard co-ordination and co-existence as shown in Fig.~\ref{fig: q=1 games} and the additional types shown in Fig.~\ref{fig: q/=1 games}. The left column shows co-existence type behaviour. The right column shows co-ordination type dynamics.

Even though mathematically we might classify a flow as co-ordination, co-existence or one of the new types, the interior fixed point(s) are often close to $x=0$ or $x=1$. The behaviour is then more like a dominance scenario. It is often insightful to plot $\dot{x}$ as a function of $x$ for specific parameters $u$, $v$ and $q$.

\begin{figure}[htbp]
    \centering
    \captionsetup{justification=justified}
    \includegraphics[scale=0.085]{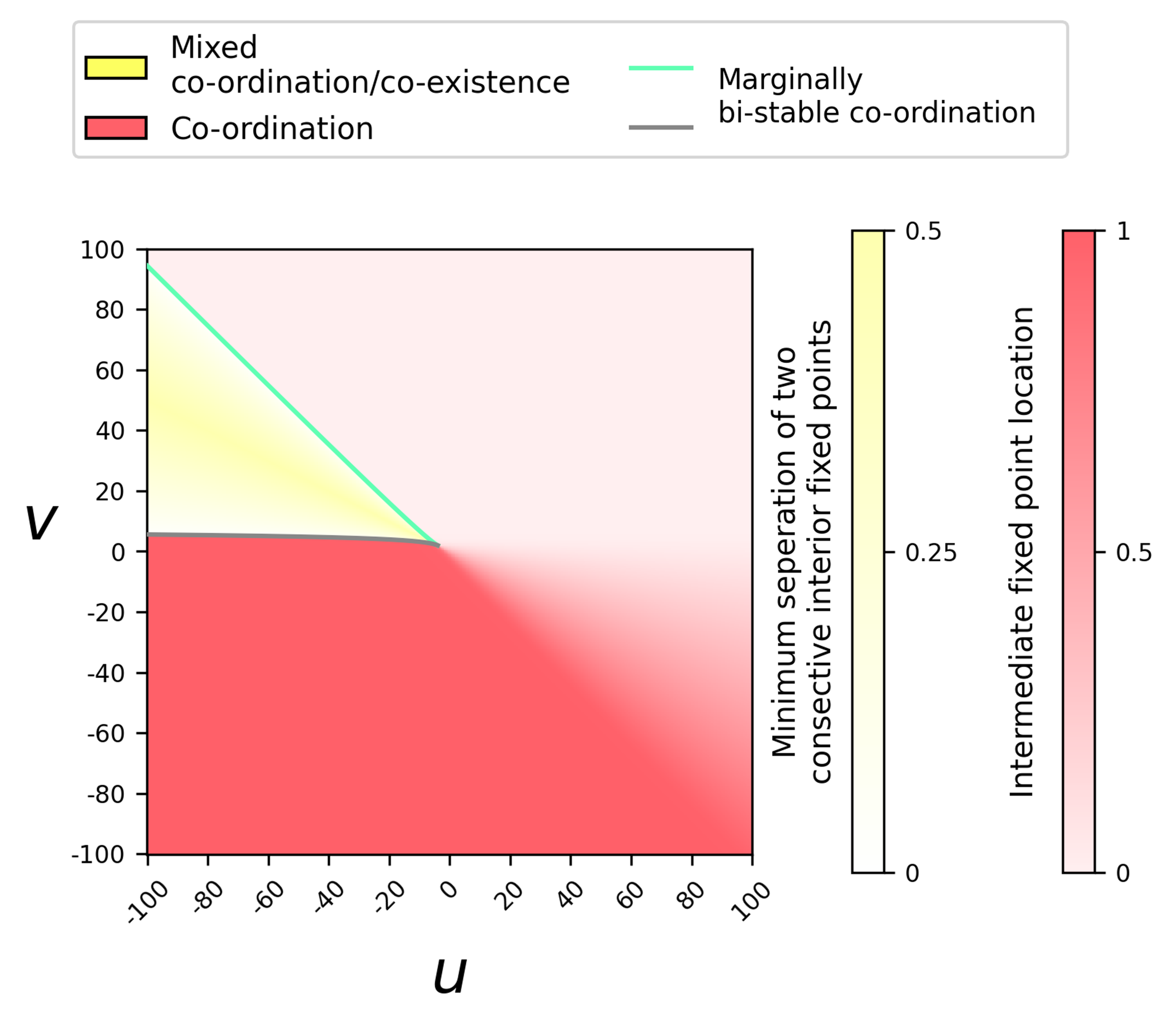}
    \caption{A phase diagram in ($u$,$v$)-space showing the classification regions for the dynamics in Eq.~(\ref{eq: q-deformed replicator}) for $q=2$, and with $g^\pm$ Fermi functions as in Eq.~(\ref{eq main: g fermi}). The opacity of the red region represents the position of the single unstable interior fixed point present in co-ordination type flows on the interval (0,1). The opacity of the yellow region represents the minimum distance between two consecutive interior fixed points in mixed co-ordination/co-existence type flows.}
    \label{fig: q/=1 phase diagram}
    \end{figure}

For the remainder of this section we focus on $q>1$ (we study the case $q<1$ in the SM, see for example Fig.~\ref{fig supp: classifying q games phase plot theory}). Figure~\ref{fig: q/=1 phase diagram} is analogous to Fig.~\ref{fig: q=1 phase diagram}, except now $q=2$. What was previously a co-existence region for $q=1$ is now bi-stable co-ordination. What was $A$/$B$-dominance for $q=1$ is now classified as co-ordination so there is one large co-ordination region where the position of the interior fixed point varies continuously. The upper and lower boundaries between the mixed co-ordination/co-existence and co-ordination regions are where marginally bi-stable co-ordination occurs.
    
\begin{figure}[htbp]
    \centering
    \captionsetup{justification=justified}
    \includegraphics[scale=0.52]{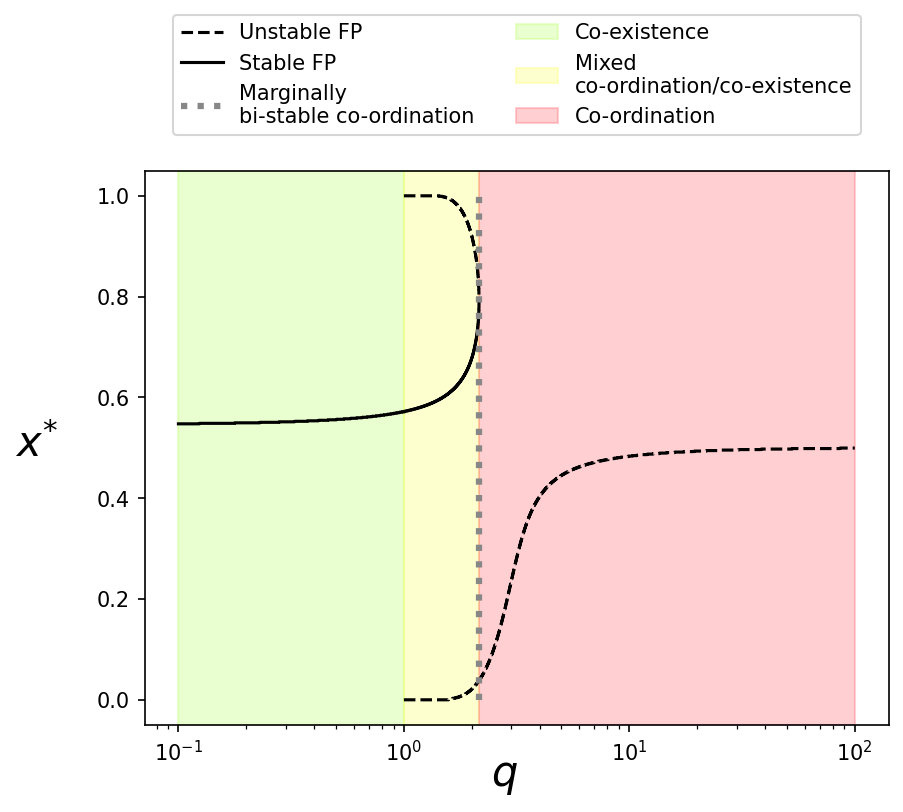}
    \caption{Bifurcation diagram showing how the interior fixed points change as the value of $q$ changes. Solid lines show stable fixed points, dashed lines indicate unstable fixed points. The figure is for $u=-7$, $v=4$. Coloured regions classify the type of game (see Fig.~\ref{fig: q=1 games}).}
    \label{fig: q/=1 bifurcation}
    \end{figure}

As we increase $q$, the shapes of these regions change. We can choose particular values of $u$ and $v$, and investigate how the fixed points move as $q$ is varied. This leads to a bifurcation diagram, as shown in Fig.~\ref{fig: q/=1 bifurcation}. For $q\leq1$ and this particular choice of $u$ and $v$ we would have a co-existence type flow, the single interior fixed point is stable. For $q>1$ we move into a mixed co-ordination/co-existence regime, and two new interior fixed points spawn at the boundaries, both of which are unstable. As $q$ increases further the stable interior fixed point and one of the unstable interior fixed points move closer to one another, and meet at $q\approx2.15$. Beyond this point, we find one interior fixed point, which is now unstable. This means that the flow is of the co-ordination type. The diagram confirms that small values of $q$ promote the minority strategy, i.e. if only a small fraction of individuals are of type $A$, then $x$ will increase, and if only a small fraction of individuals are of type $B$ then $x$ will decrease. This produces a stable internal fixed point (co-existence). Conversely, large values of $q$ promote the majority strategy, leading to co-ordination type flows.

\section{Fixation probability for a single invader, and fixation times} \label{sec: fixation times}
Having analysed the $q$-deformed deterministic dynamics in infinite populations, we now look at the fixation probabilities and times in finite populations. We will focus on the Fermi update in Eq.~(\ref{eq main: g fermi}). Our analysis follows the steps established for example in \cite{altrock2009fixation}.

The fixation probability $\phi_i$ is the probability that the evolutionary process ends in a state in which the entire population is of type  $A$ (as opposed to entirely of type $B$), if there are initially $i$ type $A$ individuals (and $N-i$ type $B$ individuals). The conditional average time to fixation $t_{1}^{A}$ is the expected time a single invader of type $A$ needs to take over a population consisting initially of $N-1$ individuals of type $B$, given such a takeover happens. The unconditional average time to fixation $t_{1}$ is the expectation value of the time until the population is monomorphic ($i=0$ or $i=N$) if there is initially one individual of type $A$ and $N-1$ individuals of type $B$. 

For the rates in Eq.~(\ref{eq: Tplusminus}), the probability $\phi_{i}$ is given by (see SM, Sec.~\ref{appendix: fixation prob})
\begin{equation}
    \phi_{i}=\frac{1+\sum_{k=1}^{i-1}f^{q-1}_{k}e^{-H_{k}}}{1+\sum_{k=1}^{N-1}f^{q-1}_{k}e^{-H_{k}}}. \label{eq main: fixation prob}
\end{equation}
Here, $f_{k}$ is defined as
\begin{equation}
    f_{k} \equiv f(k; N) = \frac{(-1)^{k}(1-N)_{k}}{k!}, \label{eq: fixation times f}
\end{equation}
where $(\cdot)_{k}$ is the Pochhammer symbol \cite{abramowitz1948handbook}. The object $H_{k}$ is defined as
\begin{equation}
    H_{k} \equiv H(k;u, v, N) = k^{2}\frac{u}{2N}+k\left(\frac{u}{2N}+v\right). \label{eq: fixation times H}
\end{equation}
The fixation times can then be written (see SM, Sec.~\ref{appendix: fixation time}) as
\begin{subequations}
\begin{align}
    t_{1} &= \phi_{1}\sum_{k=1}^{N-1}\sum_{l=1}^{k}\frac{1}{T_{l}^{+}}\left(\frac{f_{k}}{f_{l}}\right)^{q-1}e^{-(H_{k}-H_{l})},
    \label{eq: unconditional time} \\
    \nonumber \\
    t_{1}^{A} &= \sum_{k=1}^{N-1}\sum_{l=1}^{k}\frac{\phi_{l}}{T_{l}^{+}}\left(\frac{f_{k}}{f_{l}}\right)^{q-1}e^{-(H_{k}-H_{l})}. \label{eq: conditional time}
\end{align}
\end{subequations}

\begin{figure}[htbp]
    \centering
    \captionsetup{justification=justified}
    \includegraphics[scale=0.52]{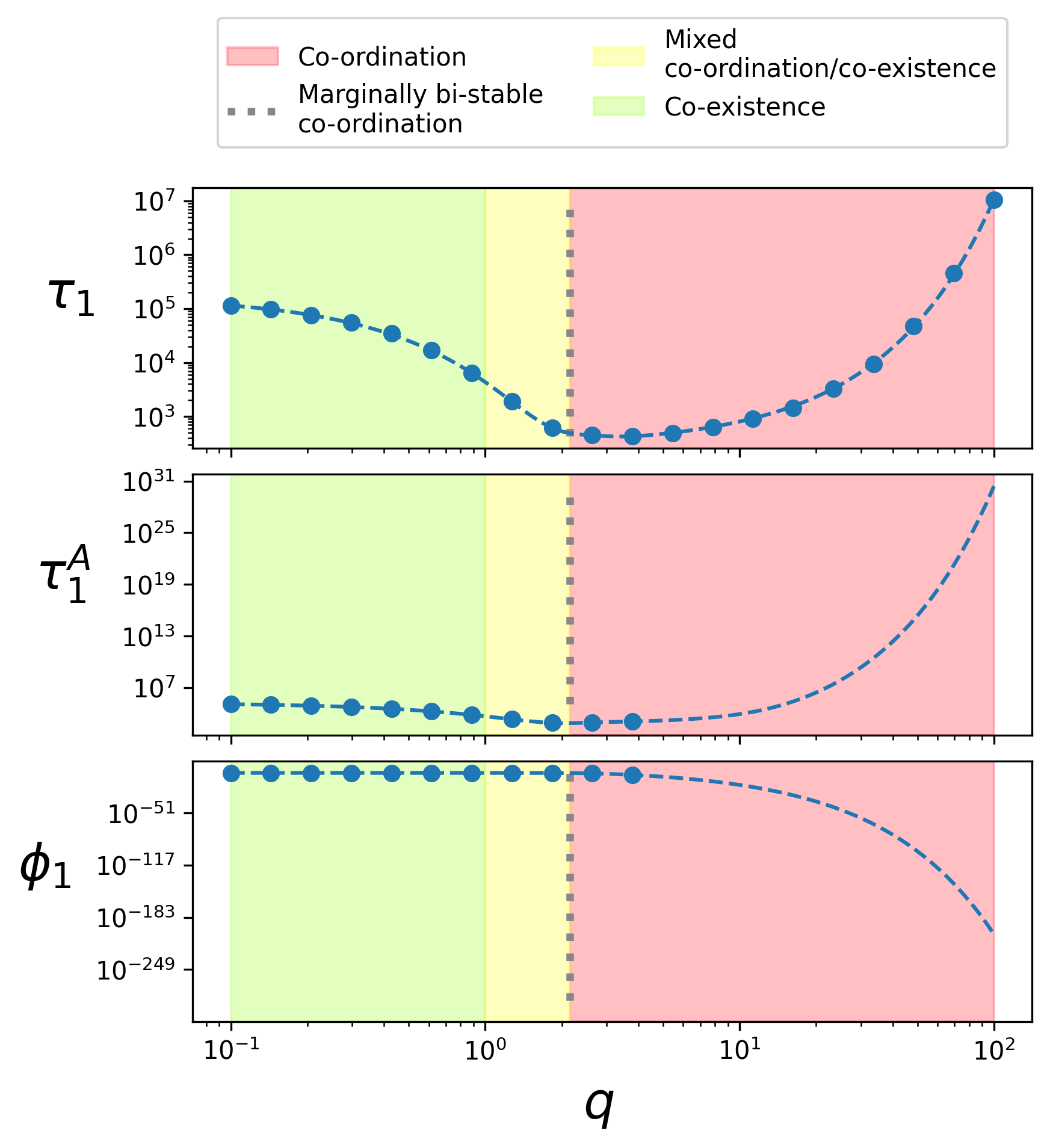}
    \caption{Conditional fixation time $t_{1}^{A}$ (upper panel), unconditional fixation time $t_{1}$ (middle panel) and fixation probability $\phi_{1}$ (lower panel) varying with $q$ for $u=-7$, $v=4$ and $N=10$ [Fig.~\ref{fig: q/=1 bifurcation} shows the corresponding bifurcation diagram for the same game (same $u$ and $v$)]. Markers are results from averaging over $1\times 10^{3}$\ independent simulations of a population evolving via the process described by Eqs.~(\ref{eq main: g fermi}) and (\ref{eq: Tplusminus}). Dashed lines are from the theory [Eqs.~(\ref{eq main: fixation prob}), (\ref{eq: unconditional time}) and (\ref{eq: conditional time})]. Coloured regions indicate the type of flow (see Fig.~\ref{fig: q=1 games}).}
    \label{fig: fixation times with q}
    \end{figure}

In Fig.~\ref{fig: q/=1 bifurcation} we showed how the interior fixed points, and hence the type of the evolutionary flow, change with $q$ for a fixed choice of $u$ and $v$. In Fig.~\ref{fig: fixation times with q} we plot the unconditional and conditional fixation times [Eqs.~(\ref{eq: unconditional time}) and (\ref{eq: conditional time})] as well as the fixation probability [Eq.~(\ref{eq main: fixation prob})] as functions of $q$, for the same game as in Fig.~\ref{fig: q/=1 bifurcation}, and for a population of size $N=10$.

For $q<1$ this particular system is of the co-existence type (green region in Figs.~\ref{fig: q/=1 bifurcation} and \ref{fig: fixation times with q}). Thus we expect trajectories of the stochastic dynamics for finite $N$ to move towards the co-existence fixed point, and then to remain near this meta-stable state until a large fluctuation drives the mutant to extinction or fixation. The escape time depends on the population size (the larger the population, the more difficult is the escape) and on the strength of the attraction of the fixed point. In the region $q<1$ the co-existence fixed points becomes less attractive as $q$ is increased (this is because for small $q$, the deformation favours the minority strategy). This is in-line with the decrease of fixation times in Fig.~\ref{fig: fixation times with q}. Consistent with this, the fixation probability decreases (from approximately $0.98$ to $0.94$ as $q$ changes from $0.1$ to $1$). This behaviour is not materially changed by the appearance of two additional unstable fixed points at $q=0$ (yellow region in Figs.~\ref{fig: q/=1 bifurcation} and \ref{fig: fixation times with q}). 

For $q\gtrsim 2.15$ the deterministic system is of the co-ordination type (red region in Figs.~\ref{fig: q/=1 bifurcation} and \ref{fig: fixation times with q}). We then find that the evolutionary flow becomes smaller in absolute terms as $q$ is increased. This flow is given by the right-hand side of Eq.~(\ref{eq: q-deformed replicator}), and describes the net change of the system per unit time. For large $q$ actual  events only occur very rarely in time, as changes in the population only happen when a sample of $q$ players all have the same type. Mathematically, the factors $x^q$ and $(1-x)^q$ become small when $x\neq 0,1$. Thus, the population dynamics becomes slower as $q$ is increased, in-line with the increase of the fixation times in the red region of Fig.~\ref{fig: fixation times with q}.

The fixation probability is not affected by time scales, but instead by the ratio of transition rates $T^-(x)/T^+(x)$, as detailed in Eq.~(\ref{eq main: fixation prob raw}) in the SM. Using Eqs.~(\ref{eq main: g fermi}) and (\ref{eq: Tplusminus}) we have 
\be
    \frac{T^{-}(x)}{T^{+}(x)} = \left(\frac{1-x}{x}\right)^{q-1}e^{-(ux+v)}.
\ee
For large $q$, the pre-factor $[(1-x)/x]^{q-1}$ dominates this expression. If $x$ is near zero the pre-factor is large, and thus, with overwhelming probability the next event in the population will lead to a decrease in mutant numbers. Similarly, when $x$ is near one, the pre-factor is small, and the next event in the population is very likely to lead to an increase of the number of mutants. In essence, $q$-deformation favours the majority type for $q>1$, and promotes co-ordination. As illustrated in Fig.~\ref{fig: gamma} in the SM this effect becomes stronger as $q$ is increased. Thus, it becomes more difficult for a single mutant to take over the population, and $\phi_1$ decreases with $q$ in the co-ordination regime, as seen in the lower panel of Fig.~\ref{fig: fixation times with q}.

\section{$q$-deformed dynamics on graphs} \label{sec: graphs}
So far we have only considered populations with all-to-all connectivity, i.e. the dynamics runs on a complete graph. We now extend the model to more general networks.

The dynamics on graphs are as follows. A random node is selected. In a second step $q$ random neighbours of that node are chosen at random with replacement. If all $q$ neighbours are in the opposite state to that of the node, it will change its state with a specific probability. This probability is as in Eq.~(\ref{eq main: g fermi}), but the payoff difference that enters into the functions $g^\pm$ is now
\be
    \Delta\pi_{n,k} = u\frac{n}{k} + v.
\ee
Here $n$ is the number of $A$ type neighbours of the node chosen for potential update, and $k$ is its degree. The quantities $u$ and $v$ are as in Eq.~(\ref{eq: uv}). Thus, $\Delta\pi_{n,k}$ is the change in the expected payoff of the focal node if it switches from $B$ to $A$. 

Using the pair approximation we can derive differential equations that can be numerically integrated to approximately describe the density of type $A$ agents, $x$, and the density of `active links', $\sigma$, on infinite uncorrelated graphs (see SM, Sec.~\ref{appendix: pair approx}). A link in the network is said to be active when it connects two individuals of different types. These equations are of the form
\BE\label{eq:pa_1}
    \frac{\dd x}{\dd t} &=& \sum_{k}P_{k}\sum_{n=0}^{k}\Bigg[(1-x)\left(\frac{n}{k}\right)^{q}g^{+}_{n,k}B_{-}(n|k) \nonumber \\
    &&-x\left(\frac{k-n}{k}\right)^{q}g^{-}_{n,k}B_{+}(n|k)\Bigg], 
\EE
and
\BE \label{eq:pa_2}
    \frac{\dd \sigma}{\dd t} &=& \frac{2}{\mu}\sum_{k}P_{k}\sum_{n=0}^{k}\Bigg[(1-x)\left(\frac{n}{k}\right)^{q}g^{+}_{n,k}B_{-}(n|k)\nonumber \\
    &&-x\left(\frac{k-n}{k}\right)^{q}g^{-}_{n,k}B_{+}(n|k)\Bigg](k-2n).  
\EE
Here $P_{k}$ is the degree distribution of the graph, and $\mu$ is the mean degree. The quantity $B_{\pm}(n|k)$ is the probability that a type $A$/$B$ node, which has degree $k$, has $n$ neighbours of type $A$. Under the pair approximation this is a binomial distribution (see SM, Sec.~\ref{appendix: pair approx}).

\begin{figure*}[htbp]
    \centering
    \captionsetup{justification=justified}
    \includegraphics[scale=0.8]{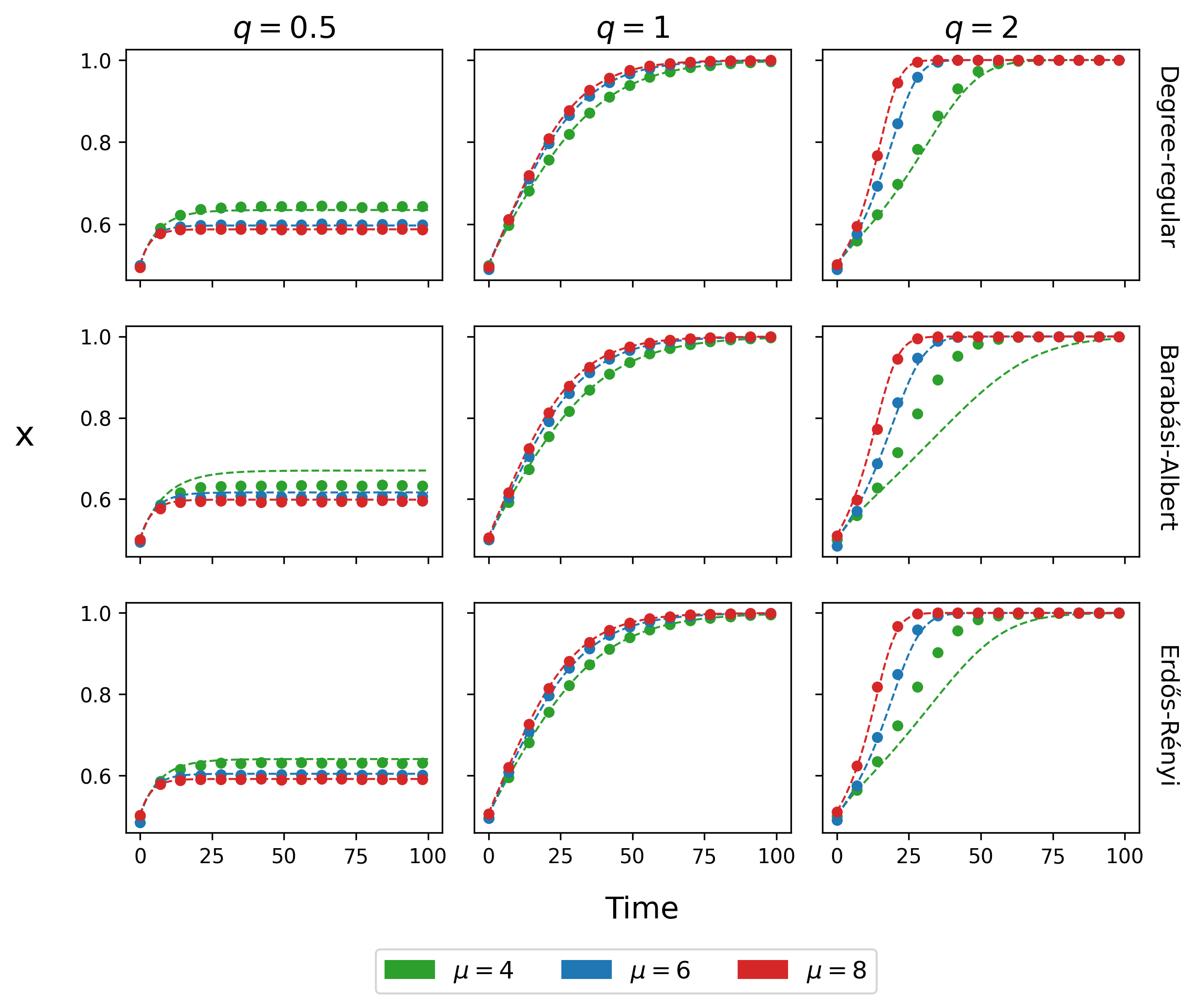}
    \caption{Density of type $A$ agents, $x$, as a function of time on various graphs with different average degree, $\mu$, for different values of $q$. Dashed lines are the analytical results from numerically integrating Eqs.~(\ref{eq supp: pair approx dx/dt final}) and (\ref{eq supp: pair approx dsigma/dt final}). Markers are results from averaging over $100$ independent trajectories. The graphs in our simulations have size $N=10,000$, and average degree as shown. The indicated average degrees can be generated exactly for degree-regular and Barab\'asi--Albert graphs but only approximately for Erd\"os--R\'enyi graphs. The game parameters are $u=0.1$ and $v=0.1$. }
    \label{fig main: graphs}
    \end{figure*}

In Fig.~\ref{fig main: graphs} we show the average fraction of agents of type $A$ as a function of time for different graphs of varying average degree and for different values of $q$. The upper three panels are for degree-regular graphs, and as the data shows the pair approximation captures simulation results well, even quantitatively. Deviations are seen for Barab\'asi--Albert and Erd\"os--Re\'nyi graphs, in particular for $q=2$ (the largest value of $q$ shown in the figure). The pair approximation correctly predicts the convergence to $x=1$, but the speed of the approach is underestimated, in particular for smaller mean degrees, where the pair approximation is known to breakdown \cite{pugliese2009heterogeneous}.

Overall, Fig.~\ref{fig main: graphs} demonstrates that the structure of the network affects the dynamics. In the examples shown, the effects of the mean degree are limited to intermediate times for $q\geq 1$ for all graphs we have tested. This is because in the long-run the system fixates $x=1$, and the stability of this state is not affected in the range of mean degrees tested in Fig.~\ref{fig main: graphs}. It is interesting to note that the dynamics in the pair-approximation can have multiple stable fixed points. We find this to be the case for $q>2$ (see Fig.~\ref{fig supp: pair approx regular flow} in the SM, indicating co-ordination-type behaviour). Convergence to $x=1$ is then only seen for some initial conditions.  

The long-term average density of type $A$ agents itself is affected for $q<1$ (left-hand panels in Fig.~\ref{fig main: graphs}). We note that the stationary density is here strictly between zero and one. Changes of the mean degree then directly alter the location of this fixed point of Eqs. (\ref{eq:pa_1}) and (\ref{eq:pa_2}).

The main purpose of this section was to describe how the pair approximation can be extended to $q$-deformed game dynamics. We stress again that the focal agent in our setup compares its current expected payoff to that it would receive if it were to change strategy. This is at variance with some existing studies of game dynamics on networks, where interaction only occurs with one single neighbour ($q=1$) \cite{szabo_fath}. If there is only one interaction partner, it is perhaps natural to compare the payoffs of the focal node and that of the interaction partner. However, for $q>1$, the focal node interacts with multiple other agents, so that payoff comparison with one single neighbour does not seem sensible. As an alternative to our dynamics one could consider a model in which the focal agent compares its current payoff to the average payoff of $q$ neighbours (but again changes can only occur provided these neighbours are all of the same type). While we expect that the pair approximation can be developed also in such a scenario, we have not pursued this here. The main reason is that the resulting theory would become more cumbersome, as the evaluation of the functions $g_{n,k}^\pm$ is then no longer local, but would also have to be based on the payoffs of the neighbours of the focal node, which in turn will depend on the degrees of those neighbours, and the states of the neighbours of the neighbours.  

\section{Multi-strategy cyclic games} \label{sec: multi strategy}
\subsection{$q$-deformed dynamics for multi-strategy games}
We now consider games with $S>2$ strategies, writing $A_{ab}$ for the payoff that an agent playing strategy $a$ receives when playing against an agent using strategy $b$. We also write  $x_a$ for the proportion of individuals in the population playing strategy $a$, and  introduce the column vector $\bx=(x_1,\dots,x_S)^T$, where $T$ stands for transpose. We have $\sum_{a=1}^S x_a=1$. The average payoff of strategy $a$ is then
\begin{equation}
    \pi_{a}(\mathbf{x}) = \sum_{b}A_{ab}x_{b}.
\end{equation}
We focus on populations with all-to-all interactions. The individual-based process is as before. We choose one individual at random for potential update, say this individual is of type $a$. In a second step $q$ individuals are sampled from the population (with replacement). Only if all of these individuals are of the same type (which we will call $b$) can a change of the original type $a$ individual occur. If this is the case, the update from $a$ to $b$ is implemented with probability $g_{a\to b}(\bx)$.

The rate for changes from $a$ to $b$ is then 
\be
    T_{a\to b} = N x_{a} x_{b}^{q}g_{a\to b}(\bx),
\ee
where we use the Fermi function and define, similar to Eq.~(\ref{eq main: g fermi}),
\begin{equation}
    g_{a\to b}(\bx) = \frac{1}{1+e^{-\beta [\pi_a(\bx)-\pi_b(\bx)]}}. \label{eq main:fermialphabeta}
\end{equation}
As in Sec.~\ref{sec: rate equations for infinite populations} we use these rates to obtain equations which govern the dynamics of the average density of type $a$ agents in infinite populations. We find
\begin{equation}
    \dot{x}_{a} = x_{a}^{q}\sum_{b\neq a}x_{b}g_{b\to a}(\bx) - x_{a}\sum_{b\neq a}x_{b}^{q}g_{a \to b}(\bx). \label{eq: N strategy q replicator}
\end{equation}
In the case of two strategies, we recover Eq.~(\ref{eq: q-deformed replicator}). 

\subsection{Three-strategy cyclic games}
As an example we consider cyclic games with three pure strategies. This generalises the well-known rock-paper-scissors (RPS) game. Following \cite{yu2016stochastic, mobilia_2010_jtb} we use the payoff matrix
\begin{equation}
    \begin{blockarray}{ccccc}
     & & \text{R} & \text{P} & \text{S} \\
    \begin{block}{cc(ccc)}
        \text{R} && 0 & -1 & 1+\delta \\
        \text{P} && 1+\delta & 0 & -1 \\
        \text{S} && -1 & 1+\delta & 0 \\
    \end{block}
    \end{blockarray}\hspace{3mm},  \label{eq main: RPS payoff}
\end{equation}
where $\delta$ is real-valued. There are then two main model parameters, $\delta$ and $q$.  Our notation follows that of \cite{yu2016stochastic}, we note that a different parametrisation is used for example in \cite{mobilia_2010_jtb}. In the SM (Sec.~\ref{appendix: multi-strategy games}) we also analyse a more general two-parameter family of payoff matrices.
\begin{figure}[hbtp]
    \centering
    \captionsetup{justification=justified}
    \includegraphics[scale=0.3]{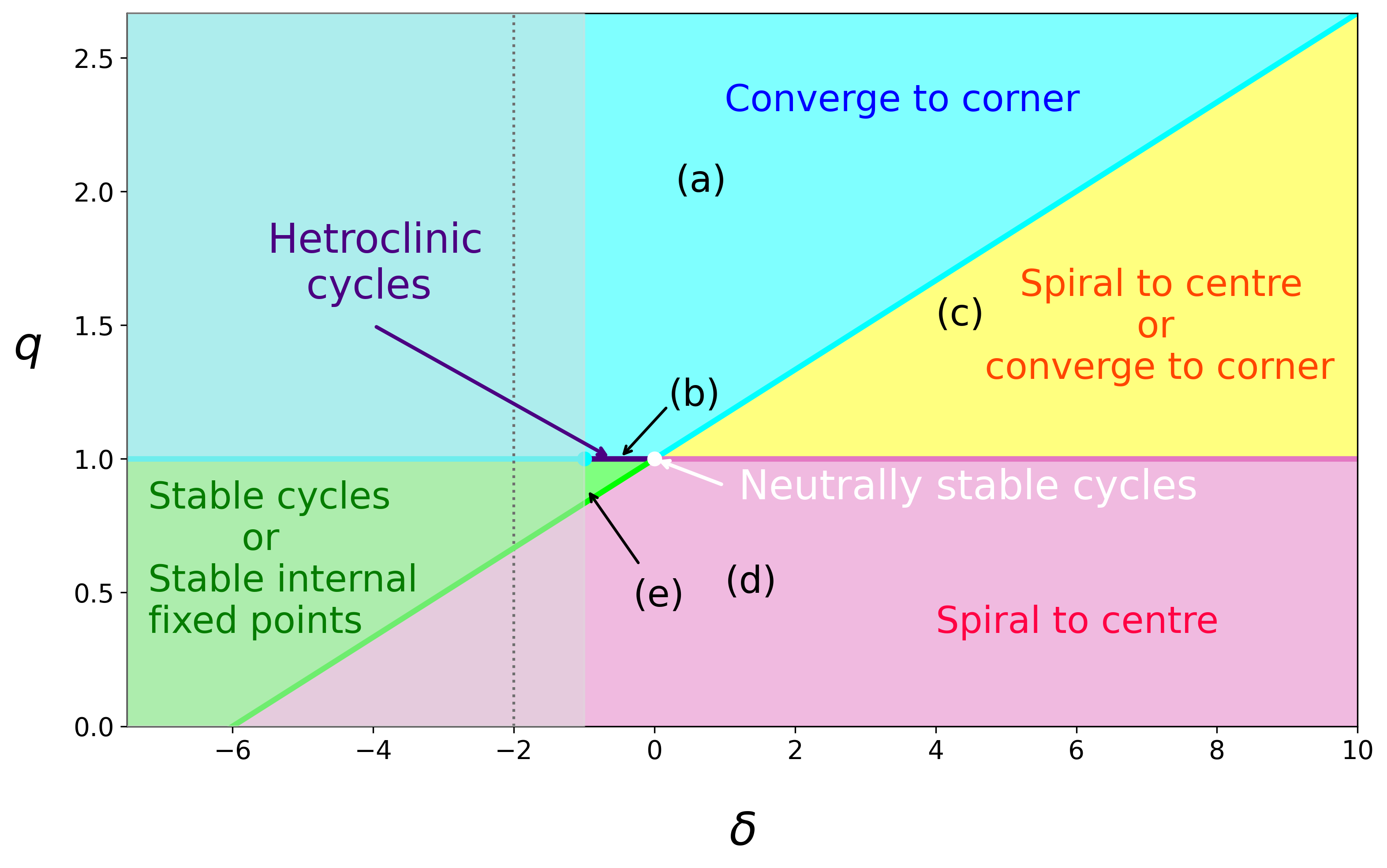}
    \caption{Types of behaviour of the deterministic rate equations for the 3-strategy cyclic game using the payoff matrix in Eq.~(\ref{eq main: RPS payoff}) as a function of $q$ and $\delta$. We mostly consider $\delta>-1$. The region $\delta\leq-1$ is also shown but greyed out. Blue: trajectories converge to one of the corners depending on initial conditions. Pink: trajectories spiral to centre point for all initial conditions. Green: trajectories converge to the same limit cycle [one for each $(q, \delta)$ pair] for all initial conditions, or trajectories can converge to stable internal fixed points within the simplex that are not the centre or corners. Indigo: trajectories converge to heteroclinic cycles on the edges of the simplex for all initial conditions. Yellow: depending on the initial conditions, trajectories will either spiral to the centre or converge to one of the corners. White: trajectories perform neutrally stable cycles for all initial conditions. By initial conditions we mean that the trajectories start from a point in the simplex that is not the centre or one of the corners. We include a feint dotted grey line at $\delta=-2$ for reference. The annotations in the plot correspond to the ternary diagrams in Fig.~\ref{fig main: simplex plots}. $(q, \delta) = [(2, 0.3), (1, -0.5), (1.5, 4), (0.5, 1), (0.87, -0.95)]$ for (a)-(e) respectively. The diagram is constructed from analysing the stability of the central fixed points and the corners, together with numerical integration of the rate equations. As a caveat we add that we cannot exclude the existence of further fixed points (see also SM, Sec.~\ref{appendix: multi-strategy games}).
    \label{fig main: q delta phase diagram}}
\end{figure}
We focus mostly on the case $\delta>-1$, such that the payoff to strategy `scissors' for example is positive when playing against `paper'. Each pure strategy then beats one other pure strategy, and is beaten by the remaining pure strategy. The case $\delta\leq -1$ can also be analysed, but the game is then not a {\em bona fide} cyclic game. 

 The centre of the strategy simplex $\left(\frac{1}{3}, \frac{1}{3}, \frac{1}{3}\right)$ is a fixed point of the dynamics for all $q$ and $\delta$, indicating co-existence of all three strategies. The monomorphic states at the corners are also fixed points (only one type of strategy survives). Performing a linear stability analysis (see SM, Sec.~\ref{appendix: multi-strategy games}) we find that the co-existence fixed point is linearly stable if and only if
\begin{equation}
    q<q_{c}(\delta)\equiv 1+\frac{\delta}{6}.
\end{equation}
The eigenvalues are complex for all $\delta\neq -1$, and thus the fixed point is a spiral sink for $q<q_c$, and a spiral source for $q>q_c$. When $q=q_{c}$ the fixed point is a centre.

A linear stability analysis of the corners show that these are stable (with two real-valued eigenvalues) for $q>1$. For $q<1$ the leading order-terms in the rate equations are sub-linear near the corners, and linear stability analysis does not apply. Nonetheless, the corners can be seen to be sources. When $q=1$ the corners are saddle points for all $\delta>-1$. Section~\ref{appendix: multi-strategy games} in the SM contains further details.

For $q=1$ the above payoff matrix is known to show different dynamics depending on the value of $\delta$ \cite{mobilia_2010_jtb,yu2016stochastic}. For $\delta > 0$ trajectories spiral to the centre of the strategy simplex, i.e. the densities of all strategies tend to $\frac{1}{3}$. When $\delta = 0$ we have neutrally stable cyclic orbits around the central fixed point. For $-1<\delta<0$ one finds heteroclinc cycles, where the trajectories orbit near the edge of the simplex. For completeness we remark that the corners of the strategy simplex are stable fixed points when $\delta\leq-1$ and $q=1$. One then finds convergence to the corners.

In Fig.~\ref{fig main: q delta phase diagram} we show a phase diagram in the $(q,\delta)$-plane highlighting the different types of outcome for the cyclic 3-strategy game for different $q$ and $\delta$. We emphasise that the behaviour in the different phases is determined from the stability analysis of the corners and the centre, combined with numerical exploration of the $q$-deformed rate equations. We cannot exclude the possibility of further fixed points in the interior, or on the edges of the simplex (see also SM, Sec.~\ref{appendix: additional ternary}). Subject to this disclaimer, we find that $q$-deformation can generate new types of flow, which we will now describe in turn.

When $\delta>0$ and $1<q<q_c(\delta)$ (yellow region in Fig.~\ref{fig main: q delta phase diagram}) the dynamics either spirals to the centre or converges to one of the corners, depending on initial conditions. For $q>\mbox{max}\{1,q_c(\delta)\}$, $q=q_{c}(\delta)$ with $\delta>0$, or $q=1$ with $\delta\leq -1$, (cyan region, including the cyan lines) we find convergence to one of the corners, which corner depends again on initial conditions. For $\delta<0$ and $q_{c}(\delta)\leq q<1$ (green region, including the green line), we either have stable limit cycles or trajectories converge to stable internal fixed points that are not the centre or the corners. The latter tends to occur for $\delta\approx -2$, an example is given in SM, Sec.~\ref{appendix: additional ternary}. In the indigo region ($q=1, -1<\delta<0$) the system shows heteroclinic cycles. For $q=1, \delta=0$ (white point in the diagram) one finds neutrally stable cycles. Finally for $q<\mbox{min}\{1,q_c(\delta)\}$, or $q=1$ with $\delta>0$, (pink region, including the pink line) the dynamics spirals to the centre. For completeness we have included some range of $\delta\leq-1$ in Fig.~\ref{fig main: q delta phase diagram}, even if the payoff matrix there does not describe a {\em bona fide} cyclic game. $\delta=-2$ is a special case where the centre is a stable/unstable star (see SM, Sec.~\ref{appendix: additional ternary}).

\begin{figure*}[htbp]
    \centering
    \captionsetup{justification=justified}
    \includegraphics[scale=0.17]{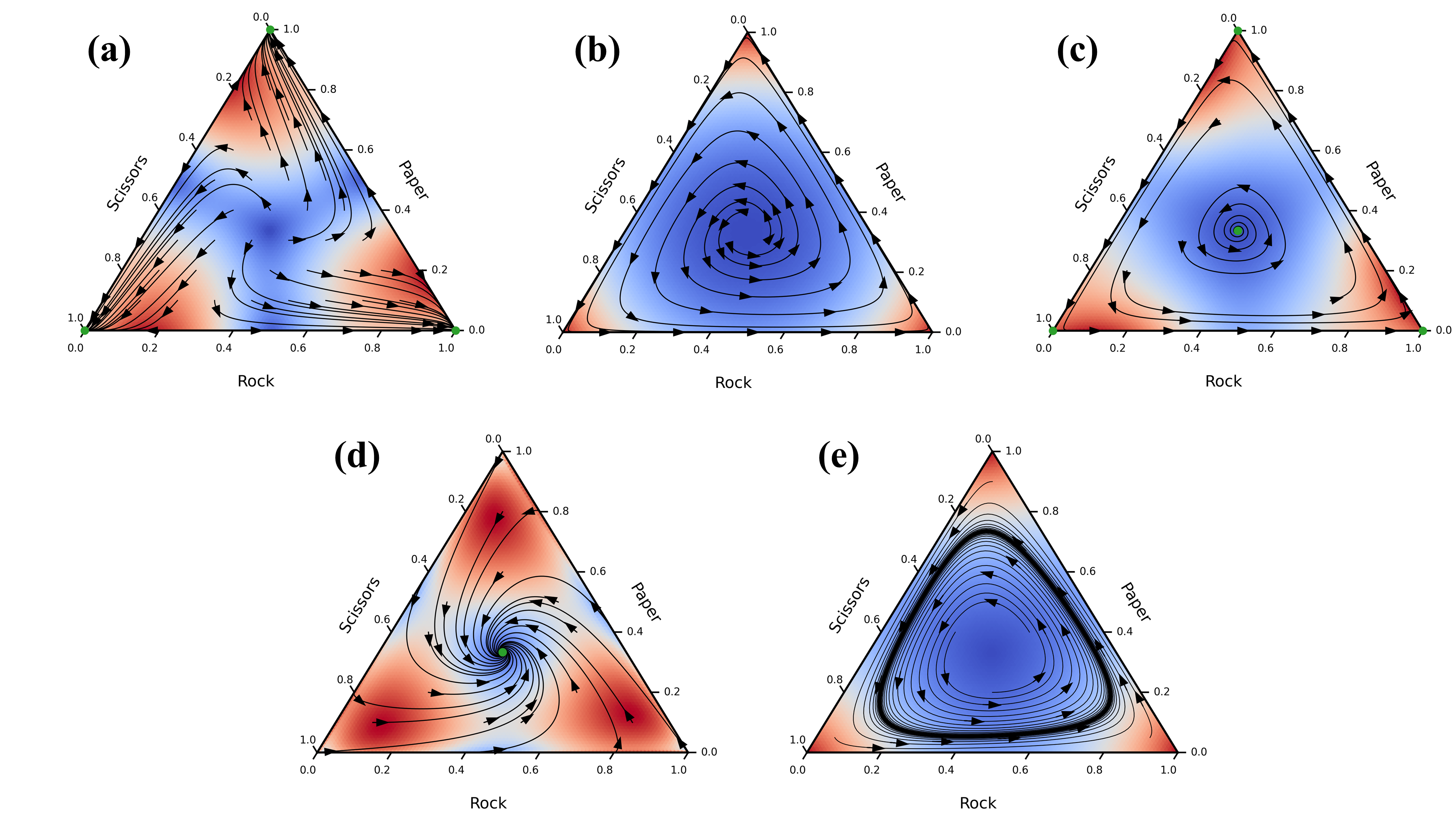}
    \caption{Ternary plots for the 3-strategy cyclic game defined by the payoff matrix in Eq.~(\ref{eq main: RPS payoff}) for different values of $(q, \delta)$ as annotated in Fig.~\ref{fig main: q delta phase diagram}. Green points are sinks. The background colour indicates the speed with which the trajectories move in the simplex, hotter colours are faster. The lines of constant $x_i$ are parallel to the tick marks on edge labelled $i\in\{\mbox{Rock, Paper, Scissors}\}$. For example, lines of constant $x_{\text{paper}}$ are horizontal.}
    \label{fig main: simplex plots}
\end{figure*}

In Fig.~\ref{fig main: simplex plots} we illustrate these different types of flow in the strategy simplex. The choice for $(q, \delta)$ are as annotated in Fig.~\ref{fig main: q delta phase diagram}.

\section{$q$-deformed dynamics without replacement} \label{sec: games without replacement}
Until now we have always assumed that the $q$ neighbours of the focal agent are chosen with replacement (i.e. the same individual can be chosen more than once).  In this section, we focus on the case without replacement. It now only makes sense to consider integer values of $q$. We focus on 2-strategy 2-player games and populations with all-to-all interaction. The state of the population is then fully described by the number $i$ of individuals playing strategy $A$.

The rates for the events $i\to i\pm 1$ in the model without replacement are
\begin{subequations}
\begin{align}
    T_{i}^{+} &= 
    \begin{cases}
    0, &0 \leq i < q \\
    (N-i)g^{+}_{i}\prod_{k=1}^{q}\left(\frac{i-k+1}{N-k}\right),  &q \leq i \leq N \\
    \end{cases} \label{eq main: T+ no rep}\\
    T_{i}^{-} &= 
    \begin{cases}
    ig^{-}_{i}\prod_{k=1}^{q}\left(\frac{N-i-k+1}{N-k}\right), &0 \leq i\leq N-q\\
    0, &N-q < i \leq N
    \end{cases} \label{eq main: T- no rep}
\end{align}
\end{subequations}
respectively, where we have defined $g^{\pm}_{i}\equiv g^{\pm}\left(i/N\right)$. We have $T^{+}_{i}=0$ for $i<q$, this is because there must be at least $q$ type $A$ agents in the system in order for a type $B$ agent to select $q$ type $A$ other agents without replacement. Similarly $T^{-}_{i}=0$ for $i>N-q$. More detail on these rates can be found in Sec.~\ref{appendix: fixation probability without replacement proof} of the SM.

\begin{figure*}[htbp]
    \centering
    \captionsetup{justification=justified}
    \includegraphics[scale=0.62]{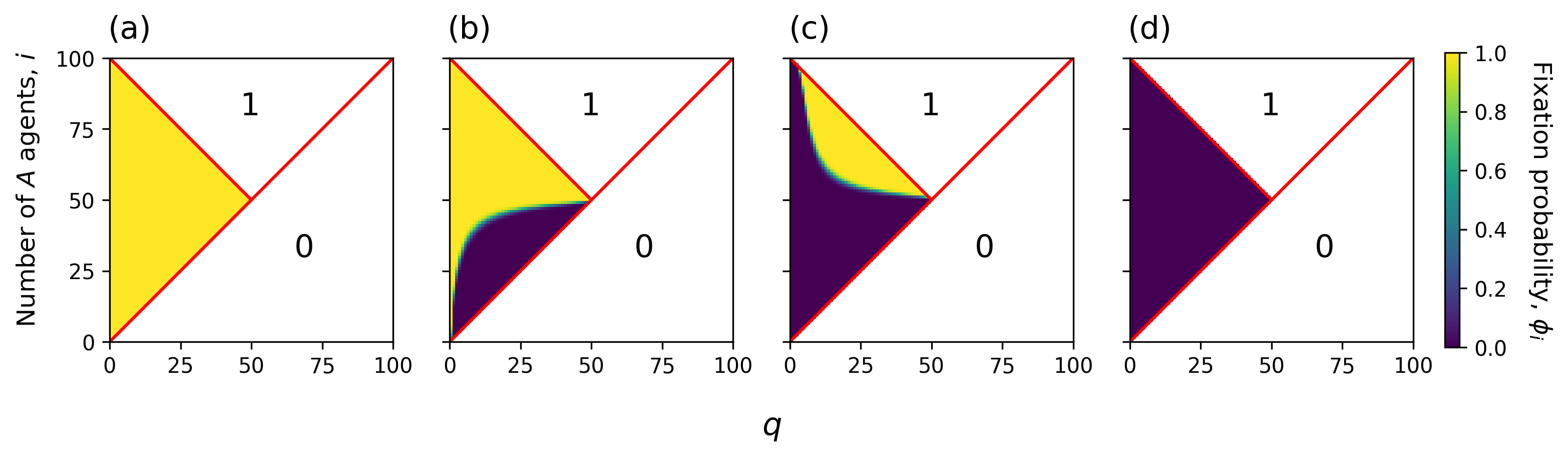}
    \caption{Fixation probability for the $q$-deformed dynamics without replacement, Eq.~(\ref{eq main: fixation prob no rep}), in the $(q,i)$-plane. This is for a population with $N=100$ for different game parameters, namely $(u, v) = [(10,10),(0.1,-0.03),(-0.1,-0.1),(-1,-1)]$ from left to right. There are two white regions, annotated with $0$ and $1$, where the fixation probability is given by those values. These regions correspond to cases one and three of Eq.~(\ref{eq main: fixation prob no rep}). In the coloured region [case two of Eq.~(\ref{eq main: fixation prob no rep})], the fixation probability is often extremely close to $0$ (purple) or $1$ (yellow), but we stress that it does not take these values exactly.}
    \label{fig main: no replacement phase plot}
\end{figure*}

Given the rates in Eqs.~(\ref{eq main: T+ no rep}) and (\ref{eq main: T- no rep}) we derive the fixation probability to be (see again Sec.~\ref{appendix: fixation probability without replacement proof} in the SM):
\begin{widetext}
\begin{equation} \label{eq main: fixation prob no rep}
    \phi_{i} = 
    \begin{cases}
        0, &(q\leq \frac{N}{2} \; \mbox{and} \; 0 \leq i < q) \; \mbox{or} \; (q>\frac{N}{2} \; \mbox{and} \; 0\leq i < q),\\
        \frac{\sum_{k=q}^{i}\prod_{j=q}^{k-1}\gamma_{j}}{\sum_{k=q}^{N-q+1}\prod_{j=q}^{k-1}\gamma_{j}}, &q \leq \frac{N}{2} \; \mbox{and} \; q\leq i \leq N-q ,\\
        1, &(q\leq \frac{N}{2} \; \mbox{and} \; N-q <i\leq N) \; \mbox{or} \; (q>\frac{N}{2} \; \mbox{and} \; q\leq i\leq N).
    \end{cases}
\end{equation}
\end{widetext}
Here $\gamma_{j}=T^{-}_{i}/T^{+}_{i}$ with $T_i^\pm$ as in Eqs.~(\ref{eq main: T+ no rep}) and (\ref{eq main: T- no rep}). The fixation probabilities are better illustrated in ($i,q$)-space, as shown in Fig.~\ref{fig main: no replacement phase plot} for different choices of $u$ and $v$. The interesting region is that defined by the middle condition of Eq.~(\ref{eq main: fixation prob no rep}) as this is the only region where the fixation probability depends on the game (in the other two regions, the fixation probability is zero or one respectively, independent of $u$ and $v$).  We highlight this in Fig.~\ref{fig main: no replacement phase plot}, the region of interest is that in the triangle on the left of each panel.

In Fig.~\ref{fig main: no replacement phase plot}(a) the fixation probability is very close to one in this area. Since $u$ and $v$ are large in panel (a) this would typically be an $A$-dominance game (see Fig.~\ref{fig: q=1 phase diagram}). At the other extreme ($u$ and $v$ sufficiently negative) the flow is mostly of the $B$-dominance type [Fig.~\ref{fig main: no replacement phase plot}(d)], and in the region of interest the fixation probability is very close to zero. For intermediate values of $u$ and $v$ the region of interest divides non-trivially into two areas, one in which $\phi_1$ is close to one, and another in which the fixation probability is close to zero [panels (b) and (c)]. 

\section{Discussion}\label{sec:conclusions}
In summary, we have combined ideas from the non-linear $q$-voter model with dynamics in evolutionary game theory. The $q$-voter model was originally introduced as a non-linear extension of the conventional voter model, and with a view towards understanding how non-linearity affects the statistical physics of simple systems with absorbing states. Evolutionary game dynamics is non-linear by itself, thus the goal of this work was to study how additional $q$-deformation affects the outcome both in infinite and in finite populations. For integer $q$ there is a clear interpretation of the dynamics: an agent in the population consults with $q$ other agents. If all those other agents are in a state which is different from that of the focal agent, that latter agent considers a change of strategy. This is then implemented with a probability based on payoff gain. Mathematically, the model can be studied for any real-valued $q$. 

As we have seen, the combination of $q$-deformation and selection because of the underlying game can produce a number of new types of flow, not seen in conventional game dynamics ($q=1$). In $2\times 2$ games multiple internal fixed points become possible, thus generating scenarios such as bi-stable co-existence or mixed co-existence and co-ordination. We have systematically studied where in the space of games these different cases occur, as $q$ is varied. In cyclic games with three strategies one also finds dynamics which are not possible without $q$-deformation, notably stable limit cycles and deterministic flow converging to pure-strategy points. We also find cases with multiple attracting fixed points in the interior of the strategy simplex (Sec.~\ref{appendix: additional ternary}). Finally, we have shown how fixation times and probabilities can be calculated for $q$-deformed dynamics in finite populations, and we have extended pair approximation methods for interacting agents on networks to $q$-deformed processes.

The main purpose of this paper is to introduce the general idea of $q$-deformation in game dynamics, to study a number of basic scenarios and to put the relevant tools in place. There are a number of lines which could be pursued in future work. For example, it might be interesting to see how $q$-deformation affects the ordering dynamics of evolutionary games on regular lattices. One might expect departures from the universality class of directed percolation for example \cite{szabo_fath}. Similarly, one could study more systematically how the system departs from the $q$-voter model in the limit of weak selection.

More generally, changing the strength of selection in an evolving population can have surprising effects. For example, so-called `stochastic slowdown' has been reported \cite{altrock_slowdown}, that is the conditional fixation time of a mutant can decrease even if the selective advantage of the mutant is increased. Non-monotonic behaviour of the conditional fixation time as a function of selection strength has also been observed \cite{altrock_slowdown_2}. Our analysis has shown that $q$-deformation also affects fixation times. It would be interesting to investigate if effects similar to stochastic slowdown also occur. We note that $q$-deformation not only changes the strength of evolutionary flow but also the location of co-existence fixed points. This is not the case in \cite{altrock_slowdown, altrock_slowdown_2}, where the position of fixed points remain the same as the strength of selection is changed.

Further aspects for future work might include multi-player games (e.g. public-good games), or asking if and how $q$-deformation affects the tendency of specific graphs to act as amplifiers or suppressors of selection \cite{liebermann,Hindersin}.

\section*{Acknowledgements}

\sloppy

This work was supported by the the Agencia Estatal de Investigaci\'on and Fondo Europeo de Desarrollo Regional (FEDER, UE) under project APASOS (PID2021-122256NB-C21, PID2021-122256NB-C22), the Mar\'ia de Maeztu programme for Units of Excellence, CEX2021-001164-M funded by  MCIN/AEI/10.13039/501100011033. We  also acknowledge a studentship by the Engineering and Physical Sciences Research Council (EPSRC, UK), reference EP/T517823/1.

\fussy


\clearpage
\newpage
\setcounter{section}{0}		
\setcounter{page}{1}		
\setcounter{equation}{0}	
\setcounter{figure}{0}		
\setcounter{table}{0}		
\renewcommand{\thesection}{S\arabic{section}} 		
\renewcommand{\thepage}{S\arabic{page}} 			
\renewcommand{\theequation}{S\arabic{equation}}  	
\renewcommand{\thefigure}{S\arabic{figure}}  		
\renewcommand{\thetable}{S\arabic{table}}  

\onecolumngrid
{\begin{center}{\Large ------ Supplemental Material ------}\end{center}}
\setlength{\parskip}{-0.6pt}
\setlength{\parindent}{0pt}
\tableofcontents

\setlength{\parskip}{8pt}
\setlength{\parindent}{0pt}
\let\addcontentsline=\oldaddcontentsline
\let\nocontentsline=\oldaddcontentsline

\FloatBarrier
\section{Deriving the \texorpdfstring{$q$}{}-deformed rate equations} \label{appendix: master eq derivation}
We focus on a well-mixed population of size $N$, and a game with two pure strategies. Define $P(x; t)$ as the probability for the system to be in state $x=i/N$ at time $t$, where $i$ is the number of type $A$ individuals. Given that the dynamics can be described via the rates $T^{+}(x)$ and $T^{-}(x)$, which increase/decrease $x$ by $\frac{1}{N}$ respectively, the master equation can be written
\begin{align}
    \frac{\dd}{\dd t}P(x; t) &= P\left(x-\frac{1}{N},t\right)T^{+}\left(x-\frac{1}{N}\right) + P\left(x+\frac{1}{N}, t\right)T^{-}\left(x+\frac{1}{N}\right) \nonumber \\
    &\qquad - P(x,t)T^{+}(x) - P(x,t)T^{-}(x). \label{eq supp: master eqn}
\end{align}
From this, the first moment $\avg{x(t)}$, which is the average density of type $A$ agents, has the following differential equation,
\begin{align}
    \frac{\dd \avg{x(t)}}{\dd t} &= \sum_{x}x\left[P\left(x-\frac{1}{N}, t\right)T^{+}\left(x-\frac{1}{N}\right)+ P\left(x+\frac{1}{N}, t\right)T^{-}\left(x+\frac{1}{N}\right)\right] \nonumber \\
    &\qquad - \sum_{x}xP(x,t)\left[T^{+}(x) + T^{-}(x)\right] \nonumber \\
    &= \sum_{x}\left(x-\frac{1}{N}\right)P(x,t)T^{+}(x) + \sum_{x}\left(x+\frac{1}{N}\right)P(x,t)T^{-}(x)\nonumber \\
    &\qquad - \avg{xT^{+}(x)} - \avg{xT^{-}(x)} \nonumber \\
    &= \avg{xT^{+}(x)} - \frac{1}{N}\avg{T^{+}(x)} + \avg{xT^{-}(x)} - \frac{1}{N}\avg{T^{-}(x)} - \avg{xT^{+}(x)} - \avg{xT^{-}(x)} \nonumber \\
    &= \frac{1}{N}\Big[\avg{T^{+}(x)} + \avg{T^{-}(x)}\Big]. \label{eq supp: master eqn intermediate}
\end{align}
Assuming the population is of infinite size, we can ignore fluctuations. This means that the probability distribution $P(x; t)$, i.e. the solution to the master equation in Eq.~(\ref{eq supp: master eqn}), is concentrated on its mean,
\begin{equation}
    P(x,t) \rightarrow P(x, t) = \delta(x-\avg{x(t)}).
\end{equation}
With this we can simplify the average rates that appear in Eq.~(\ref{eq supp: master eqn intermediate}),
\begin{equation}
    \avg{T^{\pm}(x)} = \int T^{\pm}(x)P(x, t)\dd x = \int T^{\pm}(x)\delta(x-\avg{x(t)})\dd x = T^{\pm}(\avg{x(t)}).
\end{equation}
We ease the notation by writing $\avg{x(t)}$ as just $x$, thus we have
\begin{equation}
    \frac{\dd x}{\dd t} =\frac{1}{N}\left[T^{+}(x) - T^{-}(x)\right].
\end{equation}
We can then use the rates from Eq.~(\ref{eq: Tplusminus}) to get the `$q$-deformed' dynamics in Eq.~(\ref{eq: q-deformed replicator}).

\FloatBarrier
\section{Classifying \texorpdfstring{$2\times 2$}{} \texorpdfstring{$q$}{}-deformed evolutionary dynamics} \label{appendix: number of interior fixed points}
To classify the outcome of $q$-deformed dynamics for $2\times 2$ games we determine all fixed points, $x^{*}$, and their stability.

\FloatBarrier
\subsection{Fixed points and stability}
The fixed points are obtained by setting the right-hand side of Eq.~(\ref{eq: q-deformed replicator}) to zero,
\begin{align}
     (1-x^{*})(x^*)^qg^{+}(x^{*})-x^{*}(1-x^{*})^{q}g^{-}(x^{*}) = 0. \label{eq supp: q replicator = 0}
\end{align}
Of course $x^{*}=0, 1$ are trivial solutions corresponding to fixed points on the boundaries. We will determine their stability below. To find the interior fixed points we define
\begin{equation}
    f(x) = \ln{\left(\frac{x}{1-x}\right)}-\frac{1}{q-1}\frac{g^{+}(x)}{g^{-}(x)}, \label{eq: number of fixed points f}
\end{equation}
assuming $x\neq0,1$. The solutions to $f(x^{*})=0$ would give all interior fixed points. However, this is a non-linear equation with no analytic solution in general. Despite this, it is still possible to determine the dynamics for some given function $g$.

We assume that $g$ takes the form of the Fermi function, as in Eq.~(\ref{eq main: g fermi}). Equation~(\ref{eq: number of fixed points f}) then becomes
\begin{equation}
    f_{\text{Fermi}}(x) = \ln{\left(\frac{x}{1-x}\right)}-\frac{1}{1-q}(ux+v). \label{eq supp: f fermi}
\end{equation}
This function only exists in the interval $x\in(0,1)$. We find
\begin{subequations}
\begin{align}
    \lim_{x\to 0^{+}}f_{\text{Fermi}}(x) &\to -\infty, \label{eq supp: f fermi lower bound}\\
    \lim_{x\to 1^{-}}f_{\text{Fermi}}(x) &\to +\infty. \label{eq supp: f fermi upper bound}
\end{align}
\end{subequations}
Since $f_{\text{Fermi}}(x)$ is a continuous function there must therefore be at least one zero, so there is always at least one interior fixed point. The maximum number of zeroes is determined by the number of stationary points (extrema) of $f_{\text{Fermi}}(x)$, alongside the signs of $f_{\text{Fermi}}(x)$ at those stationary points. For example, if there are two stationary points and if $f_{\text{Fermi}}(x)$ takes a positive value at one of these point, and a negative value at the other, then $f_{\text{Fermi}}(x)$ has three zeroes, and hence there are three interior fixed points. 

Differentiating Eq.~(\ref{eq supp: f fermi}) and equating to zero gives the quadratic equation
\begin{equation}
    x^{2}-x+\frac{1-q}{u} = 0, \label{eq supp: f fermi derivative}
\end{equation}
which has solutions
\begin{equation} 
    x_{1,2} = \frac{1}{2}\left[1\pm\sqrt{1-4\frac{1-q}{u}}\right]. \label{eq supp: f fermi stationary points}
\end{equation}
These are the locations of the stationary points of the function $f_{\rm Fermi}(x)$ in Eq.~(\ref{eq supp: f fermi}). Thus, depending on the values of the parameters $q$ and $u$, we can have zero, one or two stationary points.  

The possible scenarios are as follows:
\begin{enumerate}
    \item $\frac{1-q}{u}>\frac{1}{4}$, $f_{\text{Fermi}}(x)$ has no stationary points. Thus $f_{\text{Fermi}}(x)$ has one zero, meaning there is one interior fixed point.
    \item $\frac{1-q}{u}=\frac{1}{4}$, $f_{\text{Fermi}}(x)$ has one stationary point at $x^{*}=\frac{1}{2}$, so $f_{\text{Fermi}}(x)$ has one zero, and there is one interior fixed point.
    \item $0<\frac{1-q}{u}<\frac{1}{4}$, $f_{\text{Fermi}}(x)$ has two stationary points, $x_{1}$ and $x_{2}$, both of which are in the interval $x\in(0,1)$. This leads to the following sub-scenarios:
    \begin{enumerate}
        \item The two stationary points have opposite signs, i.e. $f_{\text{Fermi}}(x_{1})\cdot f_{\text{Fermi}}(x_{2}) < 0$, which means $f_{\text{Fermi}}(x)$ has three zeroes, hence there are three interior fixed points.
        \item The two stationary points have the same sign, i.e. $f_{\text{Fermi}}(x_{1})\cdot f_{\text{Fermi}}(x_{2}) > 0$, which means $f_{\text{Fermi}}(x)$ has one zero, and hence there is one interior fixed point.
        \item One stationary point lies exactly on the axis $f_{\text{Fermi}}(x)=0$, i.e. either $f_{\text{Fermi}}(x_{1})=0$ or $f_{\text{Fermi}}(x_{2}) = 0$, which means $f_{\text{Fermi}}(x)$ crosses the horizontal axis once and touches it once at another location. Hence, there are two interior fixed points. 
    \end{enumerate}
    \item $\frac{1-q}{u}\leq 0$, Eq.~(\ref{eq supp: f fermi stationary points}) has two solutions but they lie outside of the range $x\in(0,1)$. Since $f_{\text{Fermi}}(x)$ is bounded on $x\in(0,1)$ these are not valid solutions. Thus $f_{\text{Fermi}}(x)$ has no stationary points, which means it crosses the $x$-axis once, so one interior fixed point.
\end{enumerate}

Next we determine the stability of the boundary fixed points. If $q>1$, then $f_{\text{Fermi}}(x)$ and $\dot{x}=\frac{\dd x}{\dd t}$ will always be both greater than zero or both less than zero. For $q<1$ one will be greater than zero and the other less than zero. This can be easily proved by first assuming $\dot{x}>0$ [Eq.~(\ref{eq supp: q replicator = 0})]. Then manipulate the inequality into the form of Eq.~(\ref{eq supp: f fermi}), at which point we find $f_{\text{Fermi}}(x)>0$ (if $q>1$) or $f_{\text{Fermi}}(x)<0$ (if $q<1$), i.e. the same/opposite sign to $\dot{x}$. 

Now, we know from Eq.~(\ref{eq supp: f fermi lower bound}) that $f_{\text{Fermi}}(x)$ is negative as $x\to0^{+}$, thus for $q>1$, $\dot{x}$ will also be negative in this limit. This means that the fixed point at $x^{*}=0$ is stable. We can use a similar argument to show that the fixed point at $x^{*}=1$ is also stable.

The stability of the interior fixed points can be determined in the same way. For example, if $\dot{x}$ and $f_{\text{Fermi}}(x)$ have the same sign, and $f_{\text{Fermi}}(x)$ only crosses the horizontal axis [$f_{\text{Fermi}}(x)=0$] once, there is one interior fixed point, and that fixed point must be unstable. 

The above classifications are summarised in Tab.~\ref{tab: q-model classification table}. Graphical representations of these classifications are shown in Fig.~\ref{fig: q/=1 games}. We note that there are four scenarios that lead to `marginally bi-stable' flow. These types of flow are rare in comparison to the others. We group the classifications into two pairs: `marginally bi-stable co-existence' and `marginally bi-stable co-ordination'. The difference within a pair is the ordering of the two interior fixed points. 
\begin{table}
    \captionsetup{justification=justified}
    \begin{tabular}{ |p{4.5cm}|p{4.5cm}|p{1.8cm}|p{5.cm}| } 
    \hline
    \makecell{\textbf{Discriminant classification}} & \makecell{\textbf{Stationary point value}} & \makecell{\textbf{value of $q$}~~} & \makecell{\textbf{Classification}} \\
    \hline
    \multirowcell{2}{\large $\frac{1-q}{u}\geq \frac{1}{4}$ or $\frac{1-q}{u}\leq0$}& \multirowcell{2}{N/A} & \makecell{$q>1$} & Co-ordination \\ 
    \cline{3-4}
    & & \makecell{$q<1$} & Co-existence \\ 
    \hline
    \multirowcell{8}{\Large $0<\frac{1-q}{u} < \frac{1}{4}$}& \multirowcell{2}{$f_{\text{Fermi}}(x_{1})\cdot f_{\text{Fermi}}(x_{2})>0$} & \makecell{$q>1$} & Co-ordination \\ 
    \cline{3-4}
    & & \makecell{$q<1$} & Co-existence \\ 
    \cline{2-4}
     & \multirowcell{2}{$f_{\text{Fermi}}(x_{1})\cdot f_{\text{Fermi}}(x_{2})<0$} & \makecell{$q>1$} & Mixed co-ordination/co-existence \\
     \cline{3-4}
    & & \makecell{$q<1$} & Bi-stable co-existence \\ 
    \cline{2-4}
     & \multirowcell{2}{$f_{\text{Fermi}}(x_{1})=0$} & \makecell{$q>1$} & Marginally bi-stable co-ordination \\ 
     \cline{3-4}
    & & \makecell{$q<1$} & Marginally bi-stable co-existence \\ 
    \cline{2-4}
     & \multirowcell{2}{$f_{\text{Fermi}}(x_{2})=0$} & \makecell{$q>1$} & Marginally bi-stable co-ordination \\ 
     \cline{3-4}
    & & \makecell{$q<1$} & Marginally bi-stable co-existence \\ 
    \hline
    \end{tabular}
    \caption{Classification of Eq.~(\ref{eq: q-deformed replicator}) depending on the various parameters chosen. The first column evaluates the discriminant of Eq.~(\ref{eq supp: f fermi derivative}) to determine the number of stationary points of $f_{\text{Fermi}}(x)$, Eq.~(\ref{eq supp: f fermi}).  The second column evaluates $f_{\text{Fermi}}(x)$ at the stationary points, $x_{1}$ and $x_{2}$ given by Eq.~(\ref{eq supp: f fermi stationary points}), where we assume $x_{1}<x_{2}$, to determine the number of $x$-axis crosses. The third column differentiates between $q>1$ and $q<1$. The fourth column then gives the classification of the game. }
    \label{tab: q-model classification table}
\end{table}

\subsection{Phase diagram}

\FloatBarrier
\subsubsection{General analysis}
These classifications are better illustrated by a phase plot in the $(u,v)$-plane, which will also help to highlight how the $q$-deformed flow changes as we alter the payoff matrix. We first want to determine all points in the $(u,v)$-plane that would give marginally bi-stable flow. To do this we substitute the stationary point solutions $x_{1}$ and $x_{2}$, given by Eq.~(\ref{eq supp: f fermi stationary points}), into $f_{\text{Fermi}}(x)$ [Eq.~(\ref{eq supp: f fermi})], then set this equal to zero and solve. This is equivalent to solving $f_{\text{Fermi}}(x_{1,2})=0$ which, as seen from Tab.~\ref{tab: q-model classification table}, is what defines marginally bi-stable flow. We find
\begin{subequations}
\begin{align}
    v^{(1)} &= (1-q)\ln{\left(\frac{x_{1}}{1-x_{1}}\right)}-ux_{1}, \label{eq: nullcline 1}\\
    v^{(2)} &= (1-q)\ln{\left(\frac{x_{2}}{1-x_{2}}\right)}-ux_{2}. \label{eq: nullcline 2}
\end{align}
\end{subequations}
Recall from Eq.~(\ref{eq supp: f fermi stationary points}) that $x_{1,2}$ are functions of $q$ and $u$, thus for a given value of $q$, Eqs.~(\ref{eq: nullcline 1}) and (\ref{eq: nullcline 2}) are lines in the $(u,v)$-plane. These equations are only valid for $q>1$ when $-\infty < u \leq \tilde{u}$, and for $q<1$ when $\tilde{u}\leq u<\infty$, where
\begin{equation}
    \tilde{u} = 4(1-q). \label{eq supp: u star}
\end{equation}
Otherwise Eq.~(\ref{eq supp: f fermi stationary points}) does not have real solutions.

Along the line defined by $v^{(1)}$ one has $f_{\text{Fermi}}(x_{1})=0$, i.e. marginally bi-stable co-ordination and marginally bi-stable co-existence for $q>1$ and $q<1$ respectively. Similarly, sitting on the $v^{(2)}$ line corresponds to $f_{\text{Fermi}}(x_{2})=0$, again marginally bi-stable co-ordination and marginally bi-stable co-existence for $q>1$ and $q<1$ respectively. The region between these lines corresponds to $f_{\text{Fermi}}(x_{1})\cdot f_{\text{Fermi}}(x_{2})<0$, i.e. mixed co-ordination/co-existence or bi-stable co-existence for $q>1$ and $q<1$ respectively. The region outside of these lines is simply standard co-ordination or co-existence for $q>1$ and $q<1$ respectively.

\begin{figure}[htbp]
    \centering
    \captionsetup{justification=justified}
    \includegraphics[scale=0.18]{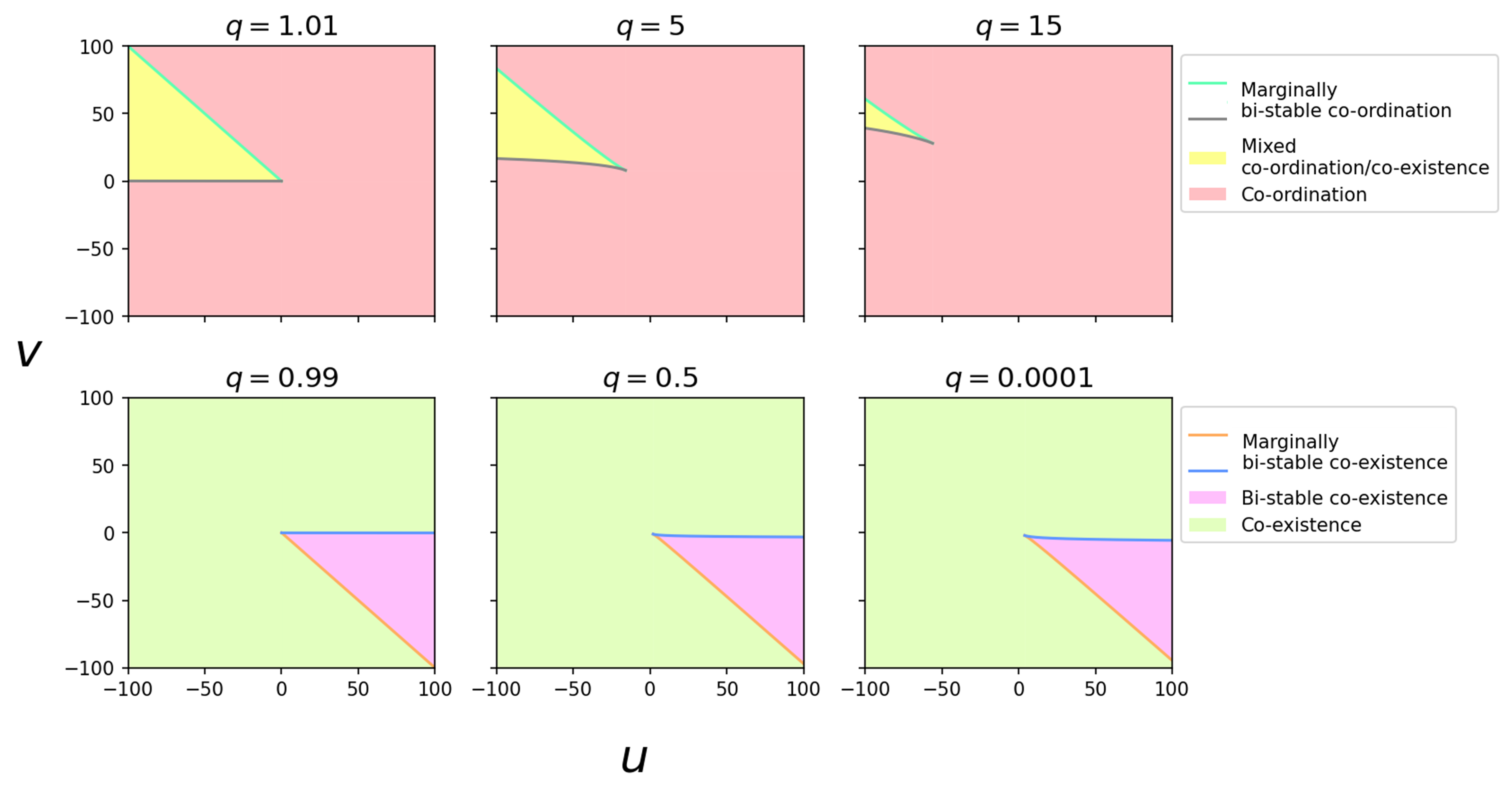}
    \caption{Phase plots in $(u,v)$-space showing how the classification regions of the $q$-deformed replicator equations change as we alter the payoff matrix for  different $q$. Regions are coloured in line with Figs.~\ref{fig: q=1 games} and \ref{fig: q/=1 games} in the main paper.}
    \label{fig supp: classifying q games phase plot theory}
    \end{figure}

Fig.~\ref{fig supp: classifying q games phase plot theory} demonstrates the idea for different values of $q$. This figure highlights the fact that marginally bi-stable co-ordination and marginally bi-stable co-existence type flow are rare, as they are only seen on well defined lines in the $(u,v)$-plane. Most of the time we see either co-ordination/co-existence, or mixed co-ordination/co-existence ($q>1$) and bi-stable co-existence ($q<1$).
    
\FloatBarrier
\subsubsection{The limit \texorpdfstring{$q\to\infty$}{}}

In the limit $q\to\infty$ both $(1-x)^{q}$ and $x^{q}$ go to zero for any $0<x<1$. Thus Eq.~(\ref{eq supp: q replicator = 0}) is satisfied for all values of $x\in(0,1)$, meaning all such points are fixed points. This is because for $q\to\infty$ no events can occur in a population containing individuals both of type $A$ and type $B$. 

\subsubsection{Large, but finite \texorpdfstring{$q$}{}}
For large but finite $q$, the mixed co-ordination/co-existence region shrinks and standard co-ordination dominates due to the fact that $\tilde{u}$ becomes large and negative [see Eq.~(\ref{eq supp: u star})], which is the upper bound for $u$ for the mixed co-ordination/co-existence region. The interior fixed point of the dominating co-ordination region is $x^{*}\approx \frac{1}{2}$ at all points in the space classified as co-ordination. This can be seen by noting that as $q$ becomes large
\begin{equation}
    f_{\text{Fermi}}(x) \approx \ln\left(\frac{x}{1-x}\right),
\end{equation}
and $\ln\left(\frac{x^{*}}{1-x^{*}}\right)=0$ has one solution, namely $x^{*}=\frac{1}{2}$. So for large positive $q$ we only get co-ordination type flow with $x^{*}\approx\frac{1}{2}$, and we conclude that $q>1$ dynamics promotes the majority, i.e. if there are less $A$ types than $B$ types the dynamics  drives the population of $A$ types down to zero. 

Since $q$ is only large, but still finite, there always exists a region of mixed co-ordination/co-existence (yellow region in Fig.~\ref{fig supp: classifying q games phase plot theory}), but only for large negative $u$, and $v>0$. This is because in the standard $q=1$ case these mixed co-ordination/co-existence regions are co-existence regions [see Fig.~\ref{fig: q=1 phase diagram}]. The dynamics promotes the minority, i.e. if there are less $A$ types the dynamics attempts to drive the number upwards. In this way, there is a balance between the the game itself favouring the minority, and the $q$-deformation for $q>1$ promoting the majority. As we increase $q$, the influence of the $q$-deformation dominates, and the mixed co-ordination/co-existence region shrinks. 

\subsubsection{The limit \texorpdfstring{$q\to 0$}{}}
In the limit $q\rightarrow 0$, the $q$-deformation favours co-existence. This time however, $\tilde{u}\rightarrow 4$ is a finite limit [see Eq.~(\ref{eq supp: u star})]. Thus there always exists a region of bi-stable co-existence (even for $q\to 0$), as seen in the lower panels of Fig.~\ref{fig supp: classifying q games phase plot theory}. The reasoning for this is analogous to that for large finite $q$: the $q$-deformed dynamics promotes the minority while the standard replicator flow ($q=1$) in this region would promote the majority [see Fig.~\ref{fig: q=1 phase diagram}]. Decreasing $q$ initially means the influence of the $q$-deformation increases, so co-existence starts to take over. However, past a certain point this effect diminishes, and decreasing $q$ further results in no change. 

\FloatBarrier
\section{Fixation probability and fixation time for \texorpdfstring{$2\times 2$}{} games with \texorpdfstring{$q$}{}-deformed dynamics} \label{appendix: fix_prob_times}

\subsection{Fixation probability} \label{appendix: fixation prob}

For an initial number of type $A$ agents, $i$, we want to calculate the fixation probability $\phi_{i}$, which is the probability for the system to reach the state $i=N$, i.e. all agents in the population are of type $A$.

It is well known that \cite{traulsen2009stochastic}
\begin{equation}
    \phi_{i} = \frac{1+\sum_{k=1}^{i-1}\prod_{j=1}^{k}\gamma_{j}}{1+\sum_{k=1}^{N-1}\prod_{j=1}^{k}\gamma_{j}}, \label{eq main: fixation prob raw}
\end{equation}
where $\gamma_{i}=\frac{T_{i}^{-}}{T_{i}^{+}}$. Using the transition rates for our model, Eq.~(\ref{eq: Tplusminus}), we can write
\begin{align}
    \gamma_{i} = \left(\frac{N-i}{i}\right)^{q-1}\frac{g^{-}_{i}}{g^{+}_{i}}.
\end{align}
We note that $g^{\pm}_{i} \equiv g^{\pm}\left(\frac{i}{N}\right)$. Using the Fermi function as the choice for $g^\pm$ [Eq.~(\ref{eq main: g fermi})], we can write the ratio
\begin{equation}
    \frac{g^{-}_{i}}{g^{+}_{i}} = e^{-\Delta \pi_{i}}.
\end{equation}
Thus we have
\begin{equation}
    \gamma_{i} = \frac{T_{i}^{-}}{T_{i}^{+}} = \left(\frac{N-i}{i}\right)^{q-1}e^{-\Delta\pi_{i}}, \label{eq: gamma}
\end{equation}
which can be interpreted as the tendency for the system to decrease the number of type $A$ agents. Again, we define $\Delta\pi_{i}\equiv \Delta\pi\left(\frac{i}{N}\right)$, where $\Delta\pi(.)$ is as in Eq.~(\ref{eq main: payoff diff}).

We now follow the lines of \cite{altrock2009fixation}. Equation~(\ref{eq main: fixation prob raw}) requires evaluating the product
\begin{align}
    \prod_{j=1}^{k}\gamma_{j} &= \prod_{j=1}^{k}\left(\frac{N-j}{j}\right)^{q-1}e^{-\Delta\pi_{j}} \nonumber \\
    &= \left(\frac{N-1}{1}\right)^{q-1}e^{-\Delta\pi_{1}}\left(\frac{N-2}{2}\right)^{q-1}e^{-\Delta\pi_{2}}...\left(\frac{N-k}{k}\right)^{q-1}e^{-\Delta\pi_{k}} \nonumber \\
    &= \left(\frac{N-1}{1}\right)^{q-1}\left(\frac{N-2}{2}\right)^{q-1}...\left(\frac{N-k}{k}\right)^{q-1}\text{exp}\left\{-\sum_{j=1}^{k}\Delta\pi_{j}\right\} \nonumber \\
    &= \left[\frac{(-1)^{k}(1-N)_{k}}{k!}\right]^{q-1}\text{exp}\left\{-\sum_{j=1}^{k}\left(u\frac{j}{N}+v\right)\right\}, \label{eq supp: fixation times gamma prod}
\end{align}
where $(\cdot)_{k}$ is the Pochhammer symbol \cite{abramowitz1948handbook} and we have used the formula for the payoff difference as in Eq.~(\ref{eq main: payoff diff}). The summation inside the exponential has a simple closed form which we will write as 
\begin{equation}
    \sum_{j=1}^{k}(u\frac{j}{N}+v)=\frac{k(k+1)}{2N}u+kv = H(k;u,v) \equiv H_{k}.
\end{equation}
We also define the following function which appears in Eq.~(\ref{eq supp: fixation times gamma prod}) as
\begin{equation}
    f_{k} \equiv f(k;N) = \frac{(-1)^{k}(1-N)_{k}}{k!}.
\end{equation}
With this we can write the fixation probability as
\begin{equation}
    \phi_{i}=\frac{1+\sum_{k=1}^{i-1}f^{q-1}_{k}e^{-H_{k}}}{1+\sum_{k=1}^{N-1}f^{q-1}_{k}e^{-H_{k}}}.
\end{equation}
In the limit $q\rightarrow 1$, we recover the known formula for the standard Fermi process (with no deformation) \cite[p; 5]{altrock2009fixation}.

\subsection{Fixation time} \label{appendix: fixation time}
We want to calculate the unconditional, $t_{1}$, and conditional, $t_{1}^{A}$, fixation times for $q$-deformed dynamics. $t_{1}^{A}$ is the time it would take type $A$ agents to take over the population, whereas $t_{1}$ is the time it would take for the type $A$ agents to take over or become extinct, both starting from a single type $A$ agent. 

For general event rates of the one-step process these can be calculated as \cite{traulsen2009stochastic}
\begin{subequations}
\begin{align}
    t_{1} &= \phi_{1}\sum_{k=1}^{N-1}\sum_{\ell=1}^{k}\frac{1}{T_{\ell}^{+}}\prod_{m=\ell+1}^{k}\gamma_{m}, \\
    t_{1}^{A} &= \sum_{k=1}^{N-1}\sum_{\ell=1}^{k}\frac{\phi_{\ell}}{T_{\ell}^{+}}\prod_{m=\ell+1}^{k}\gamma_{m},
\end{align}
\end{subequations}
where $\gamma_{k}$ is the ratio of transition probabilities, given by Eq.~(\ref{eq: gamma}), and $\phi_{\ell}$ is the probability to fixate from $l$ type $A$ agents, given by Eq.~(\ref{eq main: fixation prob}). These equations can be written in a simpler form by evaluating the product, analogous to what was done in Sec.~\ref{appendix: fixation prob},
\begin{align}
    \prod_{m=\ell+1}^{k}\gamma_{m} &= \frac{\prod_{m=\ell}^{k}\gamma_{m}}{\prod_{m=\ell}^{l}\gamma_{m}} \nonumber \\
    &= \frac{f^{q-1}_{k}e^{-\beta H_{k}}}{f^{q-1}_{\ell}e^{-\beta H_{\ell}}} \nonumber \\
    &= \left(\frac{f_{k}}{f_{\ell}}\right)^{q-1}e^{-(H_{k}-H_{\ell})}.
\end{align}
Thus the fixation times can be written,
\begin{subequations}
\begin{align}
    t_{1} &= \phi_{1}\sum_{k=1}^{N-1}\sum_{\ell=1}^{k}\frac{1}{T_{\ell}^{+}}\left(\frac{f_{k}}{f_{\ell}}\right)^{q-1}e^{-(H_{k}-H_{\ell})}, \\
    t_{1}^{A} &= \sum_{k=1}^{N-1}\sum_{\ell=1}^{k}\frac{\phi_{\ell}}{T_{\ell}^{+}}\left(\frac{f_{k}}{f_{\ell}}\right)^{q-1}e^{-(H_{k}-H_{\ell})},
\end{align}
\end{subequations}
where $f_{k}$ and $H_{k}$ are the functions defined in Eqs.~(\ref{eq: fixation times f}) and (\ref{eq: fixation times H}) respectively. 

\subsection{Further results for the system in Figs.~\ref{fig: q/=1 bifurcation} and \ref{fig: fixation times with q}}
\begin{figure}[htbp]
    \centering
    \captionsetup{justification=justified}
    \includegraphics[scale=0.8]{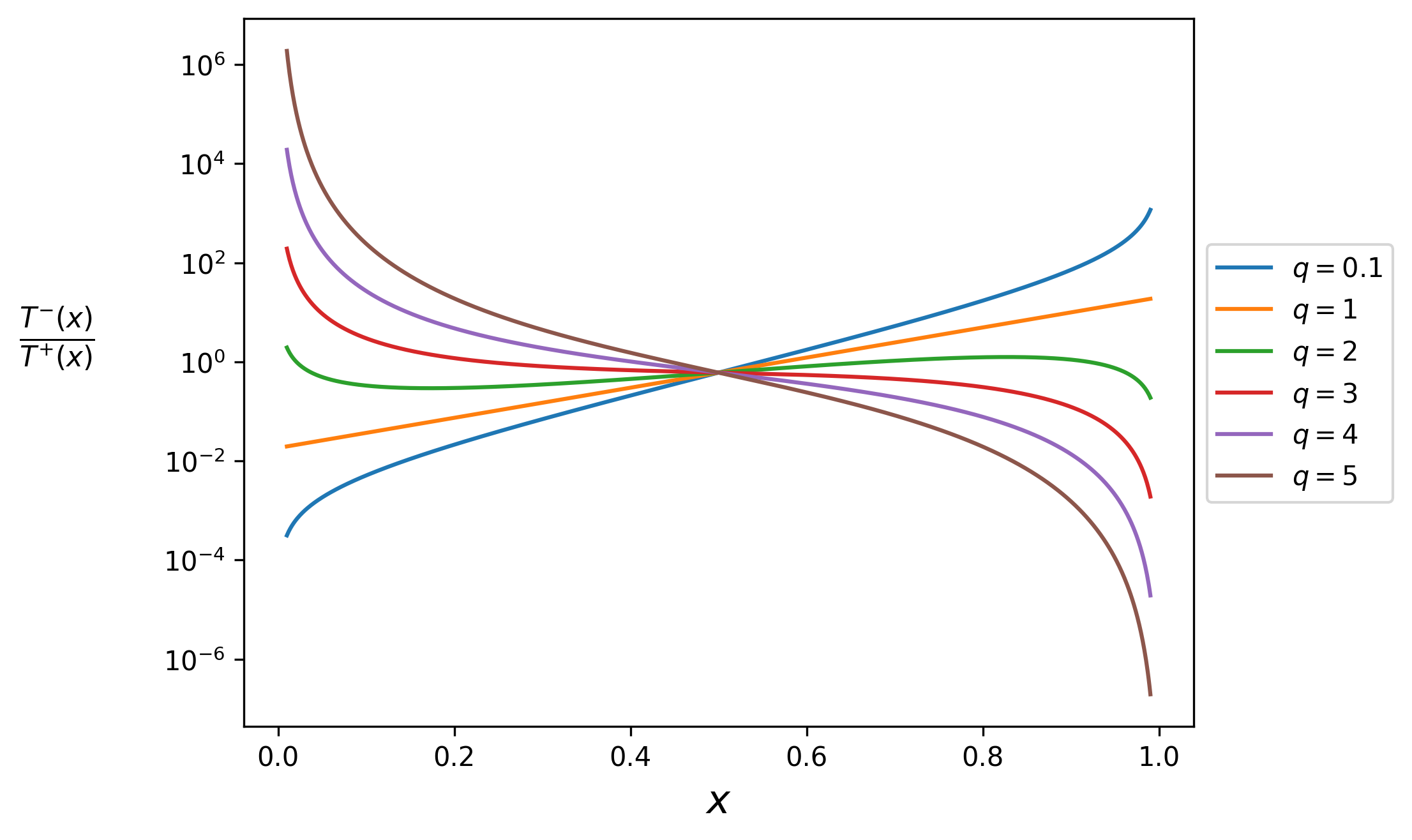}
    \caption{Ratio $T^-(x)/T^+(x)$ for the system in Figs.~\ref{fig: q/=1 bifurcation} and \ref{fig: fixation times with q}. The next event in the population increases the number of mutants (type $A$) with probability $T^+(x)/[T^+(x)+T^-(x)]=1/[1+T^-(x)/T^+(x)]$, or decreases it with probability $1/[1+T^+(x)/T^-(x)]$.}
    \label{fig: gamma}
    \end{figure}

Further results for the system in Figs.~\ref{fig: q/=1 bifurcation} and \ref{fig: fixation times with q} can be found in Fig.~\ref{fig: gamma}. For $q\gtrsim 2.15$ the flow is of the co-ordination type. The figure shows that increasing $q$ in this regime makes it more and more likely that the next event is towards the absorbing monomorphic states at $x=0$ and $x=1$. This reduces the probability for a single mutant to take over the population.
    
\section{Pair approximation on graphs} \label{appendix: pair approx}
\subsection{Notation}
We now analyse $q$-deformed dynamics on graphs. We assign each node a state $s=\pm 1$ if it is an $A$ or $B$ type respectively. We focus on undirected graphs.

We perform a \textit{homogeneous pair approximation}, similar to \cite{vazquez2008analytical, kitching2024estimating}. This approximation is known to capture the behaviour of the model to good accuracy on infinite uncorrelated graphs \cite{gleeson2013binary}. By uncorrelated graphs we mean graphs where nodes have no preference for attaching to nodes of any particular degree \cite{Dorogovtsev}. 

We write the degree distribution of a general graph as $P_{k}$. This is the probability that a randomly chosen node in the graph has degree $k$ (i.e. $k$ neighbours). We denote the number of $A$ type neighbours of a node as $n$. A node in state $s$ and of degree $k$ with $n$ type $A$ neighbours will often be referred to as an $(n,k, s)$ node.

\subsection{Rate of change of the proportion of agents of type \texorpdfstring{$A$}{}} \label{appendix: rate of change of x+}
We wish to determine a differential equation for $\avg{x_{+}}$, which we define to be the average density of nodes in the +1 state. By average we mean average over independent realisations of the dynamics. The angle-bracket notation is dropped to ease notation and keep consistency with the rest of the paper and the literature. We assume a time step of $\dd t=\frac{1}{N}$, and take $N\to\infty$ in order to obtain the continuous-time limit, thus the following now only applies to infinite graphs. 

The general rate equation for $x_{+}$ is 
\be
    \frac{\dd x_{+}}{\dd t} = \frac{1}{1/N}\sum_{s=\pm 1}T^{s}\Delta x^{s}_{+}, \label{eq supp: dxdt initial}
\ee
where  $T^{s}$ is the rate at which $-s$ nodes flip to $s$ nodes, and $\Delta x^{s}_{+}$ is the amount $x_{+}$ changes when this happens. The factor $1/N$ in the denominator results from a division by the time-step. 

It can be seen by inspection that $\Delta x^{s}_{+}=-s/N$, therefore
\be
    \frac{\dd x_{+}}{\dd t} = T^{+}-T^{-}. \label{eq supp: dx/dt}
\ee

\subsection{The homogeneous pair approximation} \label{appendix: homogeneous pair approx}
Determining the rates $T^{s}$ will require knowing the probability that a node in state $s$ with degree $k$ has $n$ neighbours in the state $+1$, we will denote this probability $B_{s}(n|k)$. To derive an expression for $B_{s}(n|k)$ we will use the \textit{homogeneous pair approximation}. This assumes that the states of the different neighbours are independent, and leads to a binomial distribution of the form
\begin{equation}
    B_{s}(n|k) = \binom{k}{n}p(+1|s)^{n}\left[1-p(+1|s)\right]^{k-n}, \label{eq supp: pair approx binom dist}
\end{equation}
where $p(+1|s)$ is the single-event probability that a node in state $s$ is connected to a node in state +1. This probability can be defined in terms of the density of state-$s$ nodes, $x_{s}$, and a quantity $\sigma$ which is the density of links connecting opposite spin nodes. We refer to $\sigma$ as the density of active interfaces/links, in-line with standard voter model terminology \cite{vazquez2008analytical}. The conditional probability $p(+|s)$ can be determined as follows: $p(+|-)$ is the ratio of the number of links connecting opposite-state nodes to the total number of $-1$ state nodes,
\be \label{eq supp: p(+|-)}
    p(+|-)=\frac{\sigma\cdot\frac{\mu N}{2}}{x_{-}\mu N} = \frac{\sigma}{2(1-x_{+})}.
\ee
The probability $p(+|+)$ can then be easily determined from 
\begin{gather}
    p(+) = p(+|-)p(-) + p(+|+)p(+), \nonumber \\
    \implies p(+|+) = 1-\frac{\sigma}{2x_{+}}. \label{eq supp: p(+|+)}
\end{gather}
Thus, combining Eqs.~(\ref{eq supp: p(+|-)}) and (\ref{eq supp: p(+|+)}) we have the general expression
\be
    p(+1|s) = \frac{1+s}{2} -s\frac{\sigma}{2x_{s}}. \label{eq supp: p(+|s)}
\ee
The moments of $B_{s}(n|k)$ can then be evaluated. As we will see below, only the first and second moments are needed, which are $\avg{n}_{s}=kp$ and $\avg{n^{2}}_{s}=k^2p^2 + kp(1-p)$ respectively, where $p$ is shorthand for $p(+1|s)$ from Eq.~(\ref{eq supp: p(+|s)}).

We emphasise that we have assumed that the probability of selecting a link connecting nodes in opposite states, $\sigma$, is independent of the degree of the nodes which it connects. This is a shortcoming of the homogeneous pair approximation. Extensions have been proposed, such as the heterogeneous pair approximation \cite{pugliese2009heterogeneous}, which allows $\sigma$ to depend on degree. Similarly, we assume an infinite graph. The stochastic pair approximation \cite{peralta2018stochastic} accounts for finite-size corrections. However, here, we do not consider these extensions.

\subsection{Rate of change of \texorpdfstring{$\sigma$}{}} \label{appendix: rate of change of sigma}
Due to the presence of $\sigma$ in the binomial moments, the differential equations for $x_{+}$ will be coupled to the differential equation for $\sigma$. The general form for such an equation has an analogous form to Eq.~(\ref{eq supp: dxdt initial}),
\be
    \frac{\dd \sigma}{\dd t} = \frac{1}{1/N}\sum_{s=\pm 1}T^{s}\Delta\sigma^s. \label{eq supp: dsigma/dt initial}
\ee
Again $T^{s}$ is the rate at which $-s$ nodes flip to $s$ nodes. The quantity $\Delta\sigma^{s}$ is the corresponding change in the density of active links. Assuming it is an $(n,k,-s)$ node that flips to a $(n,k,s)$ node, for $s=+1$ there are $n$ active links initially, and $k-n$ active links after flipping, thus a change of $k-2n$. There are $\frac{\mu N}{2}$ links overall, so the change in the density of active links is $\frac{2(k-2n)}{\mu N}$. Similar analysis for $s=-1$ gives overall $\Delta\sigma^{s}=\frac{s(k-2n)}{\mu N}$. Therefore we can write Eq.~(\ref{eq supp: dsigma/dt initial}) as
\be
    \frac{\dd \sigma}{\dd t} = \frac{2}{\mu}\left[T^{+}-T^{-}\right](k-2n). \label{eq supp: dsigma/dt}
\ee

\subsection{Transition rates on graphs}
The dynamics we have considered so far on complete graphs is a form of pairwise comparison dynamics. A node is chosen at random, $q$ of its neighbours are then chosen. If all $q$ neighbours are of opposite type to the originally selected node, the node will change type with a probability given by the Fermi function [Eq.~(\ref{eq main: g fermi})]. The argument of this function is the payoff difference $\Delta\pi$, i.e. the difference in the average payoff of the node and one of its neighbours. This is fine to do, as on a complete graph all $q$ neighbours have the same payoff (if they are all in the same state).

On general graphs things are different. A node will select $q$ random neighbours, but, even if they are all in the same state, those neighbours do not necessarily have the same average payoff, as they themselves have different neighbourhoods. Instead then on a general graph we replace $\Delta\pi\to\Delta\pi_{n,k}$ where
\begin{align}
    \Delta\pi_{n,k} &\equiv \pi_{+}(n,k) - \pi_{-}(n,k) \nonumber \\
    &= a\frac{n}{k}+b\frac{k-n}{k}-c\frac{n}{k}-d\frac{k-n}{k} \nonumber \\
    &= (a-b-c+d)\frac{n}{k}+(b-d) \nonumber \\
    &= u\frac{n}{k} + v.
\end{align}
Here $\pi_{s}(n,k)$ is the average payoff of an $(n,k,s)$ node, it is defined analogously to Eq.~(\ref{eq main: avg payoff}). $\Delta\pi_{n,k}$ then is the increment in the average payoff when an $(n,k,-1)$ node changes to an $(n,k+1)$ node. In other words, decisions are based on comparing the payoff to one node with the payoff this node would receive if it changed state (and keeping all neighbours in their present state). We thus introduce [Eq.~(\ref{eq main: g fermi})]
\begin{equation}
    g^{\pm}_{n,k}=\frac{1}{1+e^{\mp\Delta\pi_{n,k}}}. \label{eq supp: pair approx g fermi}
\end{equation}
This gives the probability that an $(n,k,-1)$ node changes to an $(n,k,+1)$ node. When $\Delta\pi_{n,k}\to\infty$, the average payoff of the node being spin $+1$ is much larger than it being spin $-1$, $g^{+}_{n,k}\to 1$ accordingly, meaning the node is guaranteed to flip.

We can now form expressions for $T^{s}$ which are needed for Eqs.~(\ref{eq supp: dx/dt}) and (\ref{eq supp: dsigma/dt}). We have
\begin{subequations}
\begin{align}
    T^{+} &= \sum_{k=1}^{N-1}P_{k}\sum_{n=0}^{k}B_{-}(n|k)x_{-}\left(\frac{n}{k}\right)^{q}g^{+}_{n,k}, \label{eq supp: graphs T+} \\
    T^{-} &= \sum_{k=1}^{N-1}P_{k}\sum_{n=0}^{k}B_{+}(n|k)x_{+}\left(\frac{k-n}{k}\right)^{q}g^{-}_{n,k}. \label{eq supp: graphs T-}
\end{align}
\end{subequations}
Consider Eq.~(\ref{eq supp: graphs T+}), which is the rate at which the density of $+1$ nodes increase under our modified pairwise comparison dynamics. For this to happen an $(n,k,-1)$ node must be picked at random, this gives rise to the first three terms. $q$ random neighbours in the state $-1$ must then picked, which happens with probability $\left(\frac{n}{k}\right)^{q}$. The node must then decide to switch its state with probability $g^{+}_{n,k}$. Eq.~(\ref{eq supp: graphs T-}) has an analogous structure where an $(n,k,+1)$ node becomes and $(n,k,-1)$ node. 

We can now substitute Eqs.~(\ref{eq supp: graphs T+}) and (\ref{eq supp: graphs T-}) into Eqs.~(\ref{eq supp: dx/dt}) and (\ref{eq supp: dsigma/dt}) to find
\begin{subequations}
\begin{gather}
    \frac{\dd x_{+}}{\dd t} = \sum_{k}P_{k}\sum_{n=0}^{k}\Bigg[x_{-}\left(\frac{n}{k}\right)^{q}g^{+}_{n,k}B_{-}(n|k)-x_{+}\left(\frac{k-n}{k}\right)^{q}g^{-}_{n,k}B_{+}(n|k)\Bigg], \label{eq supp: pair approx dx/dt final} \\
    \frac{\dd \sigma}{\dd t} = \frac{2}{\mu}\sum_{k}P_{k}\sum_{n=0}^{k}\Bigg[x_{-}\left(\frac{n}{k}\right)^{q}g^{+}_{n,k}B_{-}(n|k)-x_{+}\left(\frac{k-n}{k}\right)^{q}g^{-}_{n,k}B_{+}(n|k)\Bigg](k-2n). \label{eq supp: pair approx dsigma/dt final}
\end{gather}
\end{subequations}

\subsection{Analytically tractable limits}
We will now evaluate Eqs.~(\ref{eq supp: pair approx dx/dt final}) and (\ref{eq supp: pair approx dsigma/dt final}) in some analytically tractable limits to check their validity. These are the voter model (Sec.~\ref{appendix: voter model limit}) and the complete graph (Sec.~\ref{appendix: complete graph limit})

\subsubsection{Voter model limit} \label{appendix: voter model limit}
The limit $q=1$, $\beta=0$ (since we absorbed $\beta$ into $u$ and $v$ this means having $u=v=0$) is the standard voter model and Eqs.~(\ref{eq supp: pair approx dx/dt final}) and (\ref{eq supp: pair approx dsigma/dt final}) should reproduce known results from \cite{vazquez2008analytical}.

Eq.~(\ref{eq supp: pair approx dx/dt final}) becomes
\begin{align}
    \frac{\dd x_{+}}{\dd t} &= \frac{1}{2}\sum_{k}P_{k}\sum_{n=0}^{k}\bigg\{x_{-}\left(\frac{n}{k}\right)B_{-}(n|k)-x_{+}\left(1-\frac{n}{k}\right)B_{+}(n|k)\bigg\} \nonumber \\
    &= \frac{1}{2}\sum_{k}P_{k}\Bigg\{x_{-}\frac{\avg{n}_{-}}{k}-x_{+}\left(1-\frac{\avg{n}_{+}}{k}\right)\Bigg\} \nonumber \\
    &= \frac{1}{2}\sum_{k}P_{k}\Bigg\{x_{-}\left(\frac{\sigma}{2x_{-}}\right)-x_{+}\left(\frac{\sigma}{2x_{+}}\right)\Bigg\} \nonumber \\
    &= 0,
\end{align}
where $\avg{n}_{s}$ is the first moment of the binomial distribution $B_{s}(n|k)$ defined in Eq.~(\ref{eq supp: pair approx binom dist}). So the average density of nodes in state +1 does not change with time as expected in the standard voter model.

Eq.~(\ref{eq supp: pair approx dsigma/dt final}) becomes
\begin{align}
    \frac{\dd \sigma}{\dd t} &= \frac{1}{\mu}\sum_{k}P_{k}\sum_{n=0}^{k}\bigg\{x_{-}\left(\frac{n}{k}\right)B_{-}(n|k)-x_{+}\left(1-\frac{n}{k}\right)B_{+}(n|k)\bigg\}(k-2n) \nonumber \\
    &= \frac{1}{\mu}\sum_{k}P_{k}\Bigg\{x_{-}\left(\avg{n}_{-}-2\frac{\avg{n^{2}}_{-}}{k}\right)-x_{+}\left(k-3\avg{n}_{+}+2\frac{\avg{n^{2}}_{+}}{k}\right)\Bigg\} \nonumber \\
    &= \frac{1}{\mu}\sum_{k}P_{k}\bigg\{(k-2)\sigma - \frac{(k-1)\sigma^{2}}{2x_{+}(1-x_{+})}\bigg\} \nonumber \\
    &= \frac{1}{\mu}\Bigg\{(\mu-2)\sigma - \frac{(\mu-1)\sigma^{2}}{2x_{+}(1-x_{+})}\Bigg\}, \label{eq supp: voter model sigma diff}
\end{align}
where $\avg{n^{2}}_{s}$ is the second moment of the binomial distribution $B_{s}(n|k)$ defined in Eq.~(\ref{eq supp: pair approx binom dist}). With the replacement $x_{+}\to x$, and moving to the steady-state we find
\begin{align}
    \sigma_{\rm st} &= 2\left(\frac{\mu-2}{\mu-1}\right)x_{\rm st}(1-x_{\rm st}),
\label{eq supp: voter model sigma steady}
\end{align}
where $x_{\rm st}$ is the long-term stationary fraction of agents of type $A$. The pre-factor describes the long-lived plateau of the density of active links reported in \cite{vazquez2008analytical}.

\subsubsection{Complete graph limit}  \label{appendix: complete graph limit}
The degree distribution of a complete graph is a delta function peaked around $N$, i.e. $P_{k}\to\delta(k-N)$. Furthermore,  the binomial distributions defined in Eq.~(\ref{eq supp: pair approx binom dist}) are also delta functions peaked around $Nx_{+}$, i.e. $B_{s}(n|k)\to\delta(n-Nx_{+})$. This is because every node is connected and there are $Nx_{+}$ nodes in state +1, so the probability any node has $Nx_{+}$ neighbours is 1, and 0 for any other number of +1 neighbours. 

With this, and using the notation $x_{+}\to x$, Eq.~(\ref{eq supp: pair approx dx/dt final}) becomes
\begin{equation}
    \dot{x} = (1-x)x^{q}g^{+}(x) - x(1-x)^{q}g^{-}(x), \label{eq: pair approx rho+ diff complete graph reduction}
\end{equation}
where the standard Fermi function, Eq.~(\ref{eq main: g fermi}), has reappeared due to the delta functions acting on $g^{\pm}_{n,k}$ [Eq.~(\ref{eq supp: pair approx g fermi})]. Thus we recover Eq.~(\ref{eq: q-deformed replicator}), which applies for well-mixed populations.

Similarly, Eq.~(\ref{eq supp: pair approx dsigma/dt final}) becomes
\begin{align}
    \frac{\dd \sigma}{\dd t} &= 2(1-2x)\Big[(1-x)x^{q}g^{+}(x) - x(1-x)^{q}g^{-}(x)\Big] \nonumber \\
    &= 2(1-2x)\dot{x}, \label{eq: pair approx sigma diff complete graph reduction}
\end{align}
where we have used Eq.~(\ref{eq: pair approx rho+ diff complete graph reduction}). This is exactly as expected. As on an infinite complete graph with a fraction $x$ nodes in state +1, there are $N^{2}x(1-x)$ active links and $\frac{N^{2}}{2}$ links overall, thus $\sigma=2(x-x^{2})$. So we expected $\frac{\dd}{\dd t}\sigma = 2(1-2x)\dot{x}$, which is exactly what we see in Eq.~(\ref{eq: pair approx sigma diff complete graph reduction}).

\subsection{Degree-regular graphs} \label{appendix: regular graphs}
\begin{figure}[htbp]
    \centering
    \captionsetup{justification=justified}
    \includegraphics{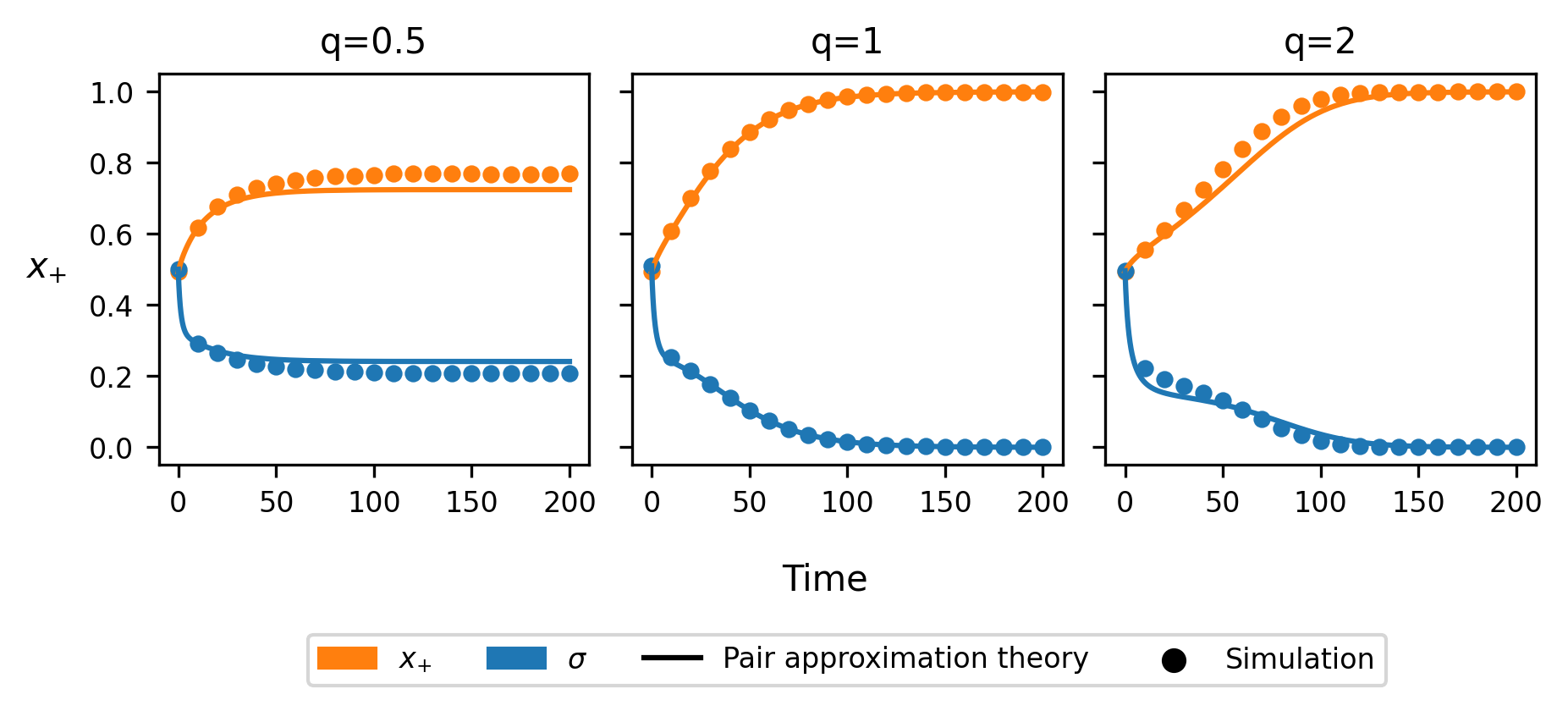}
    \caption{Plot showing the time evolution of the density of nodes in state +1 $x_{+}$ (orange), and density of active links $\sigma$ (blue) for different values of $q$, starting from a random state where $x_{+}=0.5$. The game parameters are $u=0.1$ and $v=0.1$. The underlying topology is a degree-regular graph with $N=10,000$ and $\mu=3$. Solid lines are analytical solutions from numerically integrating Eqs.~(\ref{eq supp: pair approx degree regular rho+ diff}) and (\ref{eq supp: pair approx degree regular sigma diff}). Circle markers are from averaging 100 independent Gillespie simulations of the dynamics.}
    \label{fig supp: pair approx regular graphs trajectory}
    \end{figure}

For degree-regular graphs Eqs.~(\ref{eq supp: pair approx dx/dt final}) and (\ref{eq supp: pair approx dsigma/dt final}) simplify significantly. The degree distribution is simply a delta function peaked at $\mu\in\mathbb{Z}_{\geq 2}$, i.e. $P_{k}\to\delta(k-\mu)$, which collapses the outer summations over $k$, so we have
\begin{subequations}
\begin{gather}
    \frac{\dd x_{+}}{\dd t} = \sum_{n=0}^{\mu}\Bigg[x_{-}\left(\frac{n}{\mu}\right)^{q}g^{+}_{n,\mu}B_{-}(n|\mu)-x_{+}\left(\frac{\mu-n}{\mu}\right)^{q}g^{-}_{n,\mu}B_{+}(n|\mu)\Bigg], \label{eq supp: pair approx degree regular rho+ diff} \\
    \frac{\dd \sigma}{\dd t} = \frac{2}{\mu}\sum_{n=0}^{\mu}\Bigg[x_{-}\left(\frac{n}{\mu}\right)^{q}g^{+}_{n,\mu}B_{-}(n|\mu)-x_{+}\left(\frac{\mu-n}{\mu}\right)^{q}g^{-}_{n,\mu}B_{+}(n|\mu)\Bigg](\mu-2n). \label{eq supp: pair approx degree regular sigma diff}
\end{gather}
\end{subequations}
This pair of coupled differential equations can be numerically integrated and in Fig.~\ref{fig supp: pair approx regular graphs trajectory} we show some example trajectories for a single degree-regular graph, varying the parameter $q$. The approximation works fairly well given the coarse nature of the pair approximation.

\begin{figure}[htbp]
    \centering
    \captionsetup{justification=justified}
    \includegraphics{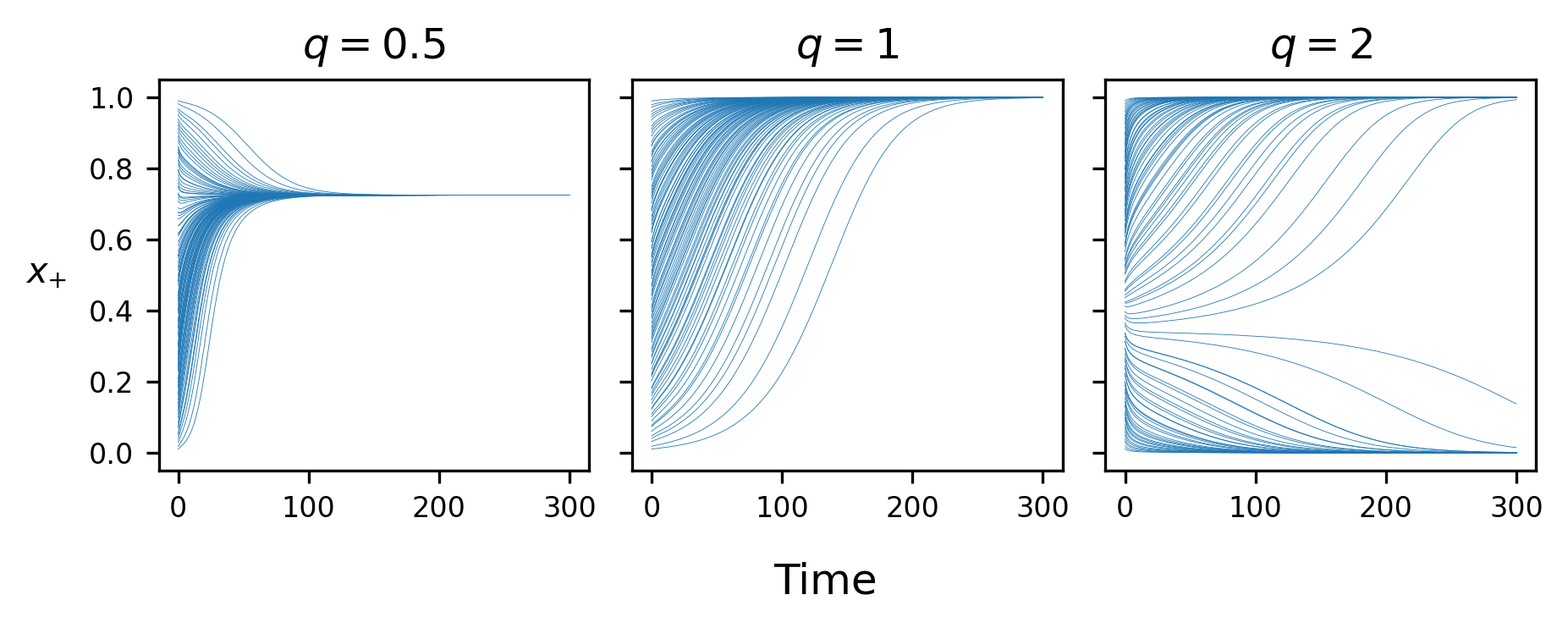}
    \caption{Pair approximation trajectories from numerically integrating Eqs.~(\ref{eq supp: pair approx degree regular rho+ diff}) and (\ref{eq supp: pair approx degree regular sigma diff}) for different values of $q$ and initial densities of $\pm 1$ type nodes. The model parameters are $u=0.1$ and $v=0.1$. The theoretical underlying degree-regular graph has $\mu = 3$ and is of infinite size.}
    \label{fig supp: pair approx regular flow}
    \end{figure}

To classify the dynamics for a given set of parameters we can look at multiple pair approximation trajectories for different initial densities of $\pm 1$ type nodes, this is shown in Fig.~\ref{fig supp: pair approx regular flow}. For $q=0.5$ there is a stable fixed point at around $x_{+}=0.7$, thus the flow could be classified as co-existence type. This is in-line with the idea that small values of $q$ promote the minority, as discussed for complete graphs in Sec.~\ref{appendix: number of interior fixed points}. For $q=1$ (standard game dynamics) there is a stable fixed point at $x_{+}=1$, which is unsurprising as the parameters chosen correspond to an $A$-dominance type flow for complete graphs [see Sec.~\ref{appendix: number of interior fixed points}]. For $q=2$ there appears to be an unstable fixed point just below $x_{+}=0.4$, thus this flow could be classified as co-ordination type. Again this is in line with Sec.~\ref{appendix: number of interior fixed points} where we found large values of $q$ promote the majority.

\FloatBarrier
\subsection{Non-regular graphs} \label{appendix: non-regular graphs}
The case of graphs which are not regular is more difficult. We use Eqs.~(\ref{eq supp: pair approx dx/dt final}) and (\ref{eq supp: pair approx dsigma/dt final}) but we have to perform the summation over the degree distribution $P_{k}$.

\begin{figure}[htbp]
    \centering
    \captionsetup{justification=justified}
    \includegraphics{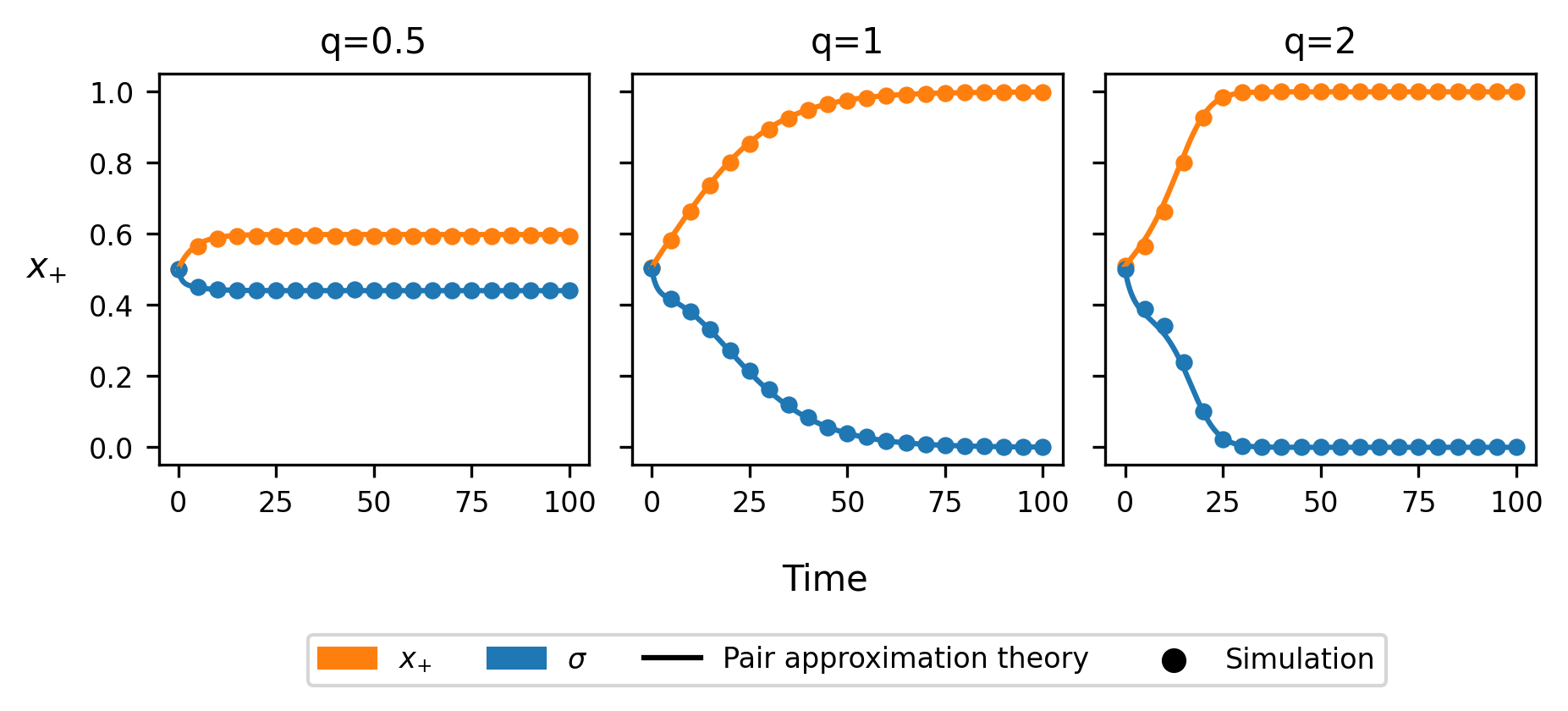}
    \caption{Plot showing the time evolution of the density of nodes in state +1 $x_{+}$ (orange), and density of active links $\sigma$ (blue) for different values of $q$, starting from a random state where $x_{+}=0.5$. The game parameters are $u=0.1$ and $v=0.1$. The underlying topology is a Barab\'asi--Albert graph with $N=10,000$ where each node connects to 4 existing nodes so that $\mu = 8$. Solid lines are analytical solutions from numerically integrating Eqs.~(\ref{eq supp: pair approx dx/dt final}) and (\ref{eq supp: pair approx dsigma/dt final}). Circle markers are from averaging 100 independent Gillespie simulations of the dynamics.}
    \label{fig supp: pair approx barabasi trajectory}
    \end{figure}

\begin{figure}[htbp]
    \centering
    \captionsetup{justification=justified}
    \includegraphics{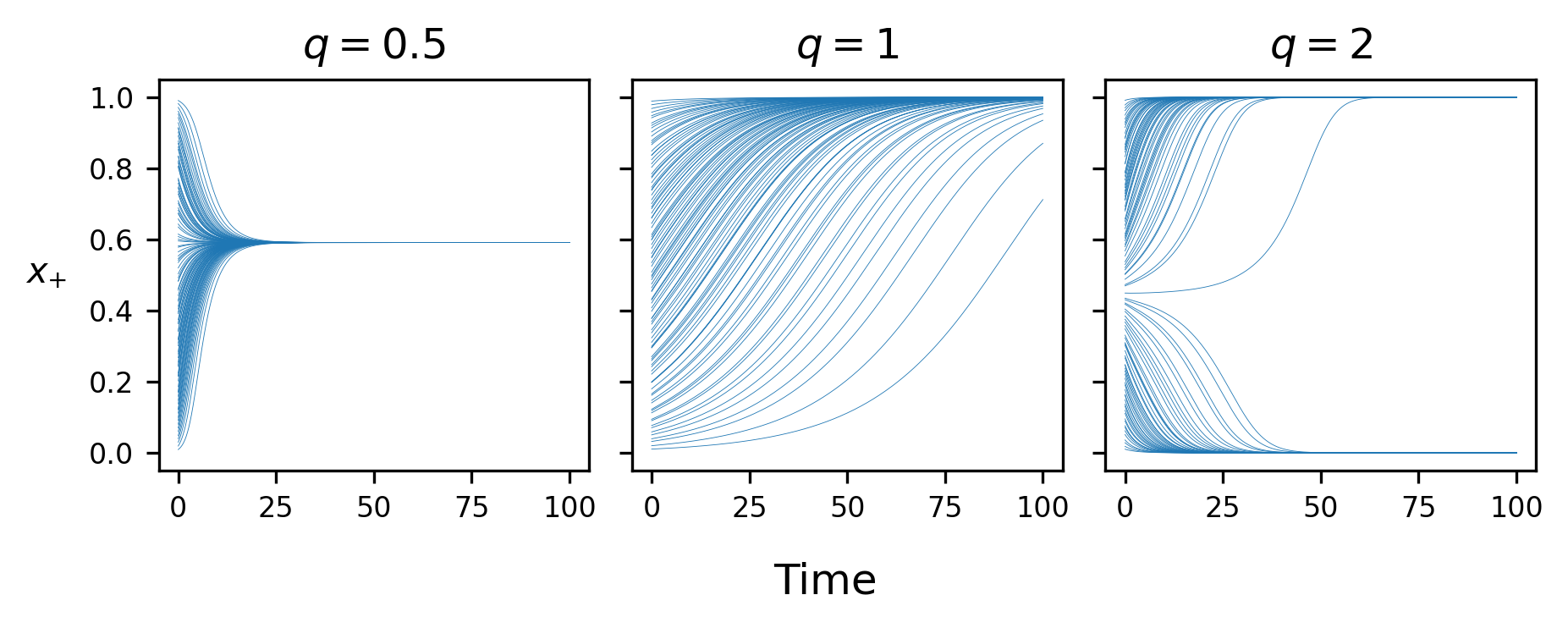}
    \caption{Pair approximation trajectories from numerically integrating Eqs.~(\ref{eq supp: pair approx dx/dt final}) and (\ref{eq supp: pair approx dsigma/dt final}) for different values of $q$ and initial densities of $\pm 1$ type nodes. The model parameters are $u=0.1$ and $v=0.1$. The underlying Erd\"os--R\'enyi graph has $N=10,000$ and $p=8/N$ so that $\mu\approx 8$.}
    \label{fig supp: pair approx erdos flow}
    \end{figure}

We show in Figs.~\ref{fig supp: pair approx barabasi trajectory} and \ref{fig supp: pair approx erdos flow} analogous plots to Figs.~\ref{fig supp: pair approx regular graphs trajectory} and \ref{fig supp: pair approx regular flow} but now for Barab\'asi--Albert and Erd\"os--R\'enyi graphs respectively to demonstrate the validity of the solution for different heterogeneous uncorrelated (or approximately uncorrelated) graphs.

We note that in Fig.~\ref{fig supp: pair approx barabasi trajectory} the analytical results agree very well with simulation, even more than for degree-regular graphs (Fig.~\ref{fig supp: pair approx regular graphs trajectory}). This is likely because in Fig.~\ref{fig supp: pair approx regular graphs trajectory} $\mu=3$ whereas in Fig.~\ref{fig supp: pair approx barabasi trajectory} $\mu=8$. This form of pair approximation is known to breakdown at lower $\mu$ \cite{pugliese2009heterogeneous}.

\FloatBarrier
\section{\texorpdfstring{$q$}{}-deformed dynamics for cyclic games} \label{appendix: multi-strategy games}

We wish to determine the classification of the $q$-deformed dynamics for cyclic games [such as the rock-paper-scissors (RPS) game]. In this appendix we study games defined by the payoff matrix
\begin{equation}
    \begin{blockarray}{ccccc}
     & & \text{R} & \text{P} & \text{S} \\
    \begin{block}{cc(ccc)}
        \text{R} && 0 & a & b \\
        \text{P} && b & 0 & a \\
        \text{S} && a & b & 0 \\
    \end{block}
    \end{blockarray}\hspace{3mm}.  \label{eq app: RPS payoff}
\end{equation}
This is a generalisation of Eq.~(\ref{eq main: RPS payoff}). We study the dynamics for general $q\in \mathbb{R}^{+}$ and $a,b\in\mathbb{R}$. 

The centre of the strategy simplex $\left(\frac{1}{3},\frac{1}{3},\frac{1}{3}\right)$ is always a fixed point, as are the pure strategies at the corners of the simplex. We thus study the linear stability of these fixed points. We note that there are only two degrees of freedom in the system since $\sum_{i}x_{i}=1$.

An accompanying Mathematica notebook with further details of the following calculations can be found at a GitHub repository \cite{github}.

\subsection{The centre point} \label{appendix: centre}
Here we analyse the stability of the point $p_{\text{centre}}=\left(\frac{1}{3},\frac{1}{3},\frac{1}{3}\right)$. The Jacobian of the dynamics in Eq.~(\ref{eq: N strategy q replicator}) (after reduction to two degrees of freedom) evaluated at the centre is 
\begin{equation}
    \renewcommand{\arraystretch}{2}
    \mathbf{J}_{\text{centre}} = \hspace{2mm}
    \begin{blockarray}{(cc)}
        \frac{1}{2\cdot 3^{q}}\left(3q-3-b\right) & \frac{1}{2\cdot 3^{q}}(a-b) \\
        -\frac{1}{2\cdot 3^{q}}(a-b) & \frac{1}{2\cdot 3^{q}}\left(3q-3-a\right) \\
    \end{blockarray}\hspace{2mm}.
\end{equation}
The eigenvalues of this matrix are,
\be
    \lambda^{\pm} = \frac{1}{4\cdot 3^{q}}\left[6q-6-(a+b)\pm i\abs{a-b}\sqrt{3}\right].
\ee
Assuming $a\neq b$, both eigenvalues are complex, and thus the point $p_{\text{centre}}$ can only be a centre, spiral sink, or spiral source \cite{strogatz2018nonlinear}. For centres, we require the real part of the eigenvalues to be zero, this occurs when
\begin{equation}
    q=q_{c}\equiv 1+\frac{a+b}{6}. \label{eq supp: centre critical q}
\end{equation}
When $q<q_{c}$, the real part of both eigenvalues is negative, so $p_{\text{centre}}$ will be a spiral sink. When $q>q_{c}$, the real part of both eigenvalues is positive, so $p_{\text{centre}}$ will be a spiral source.

In the special case of $a=b$, we have a single real eigenvalue with algebraic multiplicity 2,
\be
    \lambda  = \frac{1}{2\cdot 3^q}(3q-3-a).
\ee
The eigenspace has geometric multiplicity 2, i.e. there are two linearly independent eigenvectors corresponding to this eigenvalue, namely $\icol{1 \\ 0}$ and $\icol{0 \\ 1}$. This means that $p_{\text{centre}}$ is a star, which is a source when $\lambda>0$, i.e. $q>1+\frac{a}{3}$, and a sink when $\lambda<0$, i.e. $q<1+\frac{a}{3}$. When $q=1+\frac{a}{3}$ we have a centre once again.

For general $a$ and $b$ then, $p_{\text{centre}}$ is some form of sink when $q<q_{c}$, some form of source when $q>q_{c}$, and a centre when $q=q_{c}$.

\subsection{The corners} \label{appendix: corners}
Here we analyse the stability of the corners, $p_{\text{corner}}$, i.e. permutations of the point $\left(1,0,0 \right)$. By symmetry all corners give the same results from stability analysis.

\subsubsection{\texorpdfstring{$q = 1$}{}}
When $q=1$ the eigenvalues of $\mathbf{J}_{\text{corner}}$ are
\begin{subequations}
    \begin{gather}
        \lambda_{1} = \Tanh{\frac{a}{2}}, \\
        \lambda_{2} = \Tanh{\frac{b}{2}}.
    \end{gather}
\end{subequations}
Thus both eigenvalues are real. If both $a,b>0$, then the corners are \textbf{sources}. If both $a,b<0$, then the corners are \textbf{sinks}. If $a$ and $b$ have opposite signs then the corners are \textbf{saddle points}.

For $a=b=0$ all entries in the payoff matrix are zero. There is thus no actual game dynamics, and the flow is solely determined by the $q$-deformation. For the case $q=1$ (no deformation), which we are discussing here, this means that there is no dynamics at all ($\dot x_i=0$ for all $i$).

\subsubsection{\texorpdfstring{$q>1$}{}}
When $q>1$ the eigenvalues of $\mathbf{J}_{\text{corner}}$ are
\begin{subequations}
    \begin{gather}
        \lambda_{1} = \frac{-1}{1+e^a}, \\
        \lambda_{2} = \frac{-1}{1+e^b}.
    \end{gather}
\end{subequations}
Thus for all $a,b$ both eigenvalues are real and negative, hence the corners are always \textbf{sinks}.

\subsubsection{\texorpdfstring{$0<q<1$}{}}
For any $q$ the rate equations are (after reduction to two degrees of freedom, which we will call $x$ and $y$, these are the proportions of two of the strategies):
\begin{subequations}
    \begin{align}
        \frac{\dd x}{\dd t} &= x^{q}\left[\frac{1-x-y}{1+\text{exp}\left\{a(x-y)-b(1-x-2y)\right\}}+\frac{y}{1+\text{exp}\left\{a(1-x-2y)-b(1-2x-y)\right\}}\right] \nonumber \\
        &\qquad -x\left[\frac{(1-x-y)^{q})}{1+\text{exp}\left\{-a(x-y)+b(1-x-2y)\right\}}+\frac{y^q}{1+\text{exp}\left\{-a(1-x-2y)+b(1-2x-y)\right\}}\right], \label{eq supp: corner rate 1} \\
        \frac{\dd y}{\dd t} &= y^{q}\left[\frac{1-x-y}{1+\text{exp}\left\{-a(1-2x-y)-b(x-y)\right\}}+\frac{x}{1+\text{exp}\left\{-a(1-x-2y)+b(1-2x-y)\right\}}\right] \nonumber \\
        &\qquad -y\left[\frac{(1-x-y)^{q})}{1+\text{exp}\left\{a(1-2x-y)+b(x-y)\right\}}+\frac{x^q}{1+\text{exp}\left\{a(1-x-2y)-b(1-2x-y)\right\}}\right]. \label{eq supp: corner rate 2}
    \end{align}
\end{subequations}
The right-hand sides evaluate to zero at the corners of the strategy simplex ($x=y=0$, $x=1, y=0$, and $x=0, y=1$ respectively). Due to the symmetry with respect to interchange of types, it is sufficient to study the stability of the fixed point at one of the three corners, here we choose $x=y=0$. For $x$ and $y$ small, but non-zero, and keeping in mind that we focus on $0<q<1$, the leading order terms on the right are proportional to $x^q$ or $y^q$. Therefore, linear stability analysis cannot be used. When both $x$ and $y$ are small, Eqs.~(\ref{eq supp: corner rate 1}) and (\ref{eq supp: corner rate 2}) reduce to 
\begin{subequations}
    \begin{align}
        \frac{\dd x}{\dd t} &= \frac{x^{q}}{1+e^{-b}}, \\
        \frac{\dd y}{\dd t} &= \frac{y^{q}}{1+e^{-a}}.
    \end{align}
\end{subequations}
Therefore, both $\dot{x}, \dot{y} > 0$ for $x, y$ small (but positive). This is the case for all $a$ and $b$, so the corners are always \textbf{sources}.

\subsection{Reduction to one-parameter family of payoff matrices}
We have shown in Sec.~\ref{appendix: corners} that the stability of the corners is independent of $a$ and $b$, except when $q=1$ where only the signs of $a$ and $b$ matter. In Sec.~\ref{appendix: centre} we showed that $a+b$ determines the stability of the centre.

In principle our analysis allows for a complete classification of the flows (as far as linear stability analysis goes) for all values of $a, b$ and $q$.

In order to reduce the number of parameters we now fix $a=-1$ and focus on positive values of $b$. This reflects the scenario of a cyclic game. Each strategy is beaten by one other strategy, and in turns beats the remaining strategy.  It is convenient to set $b=1+\delta$ (following \cite{yu2016stochastic}). In this way $\delta=0$ reflects the standard zero-sum rock-paper-scissors game. The constraint $b>0$ translates into $\delta>-1$. We note that other (equivalent) setups have been used, see for example \cite{mobilia_2010_jtb}. For this reduced setup Eq.~(\ref{eq supp: centre critical q}) becomes
\be
    q_{c} (\delta) = 1+\frac{\delta}{6}. 
\ee
The centre fixed point is a stable spiral for $q<q_c(\delta)$ and an unstable spiral for $q>q_c(\delta)$. It is a centre for $q=q_c(\delta)$.

The corners are sources for $q<1$ and sinks for $q>1$, as this is independent of $\delta$. For $q=1$ the corners are saddle points for all $\delta>-1$, otherwise they are sinks.

\subsection{Further ternary plots} \label{appendix: additional ternary}
As mentioned in Sec.~\ref{appendix: centre}, when $a=b$ (or in the reduced setup $\delta=-2$) the centre point is a stable or unstable star depending on the value of $q$. 

\begin{figure}[htbp]
    \centering
    \captionsetup{justification=justified}
    \includegraphics[scale=0.16]{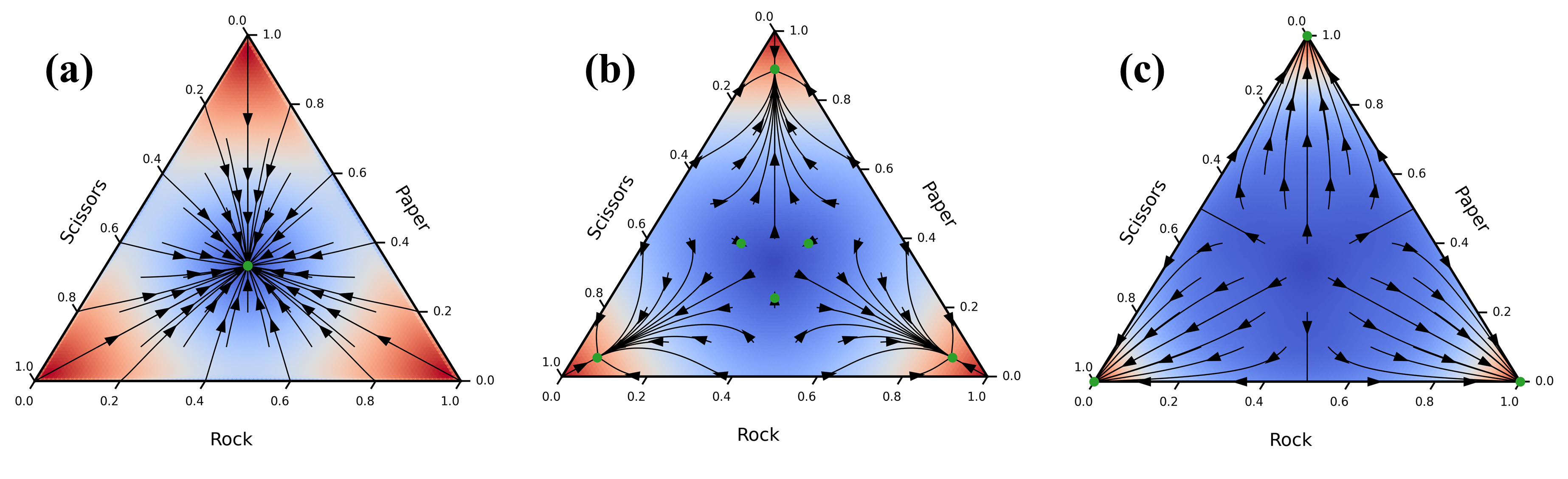}
    \caption{Ternary plots for the 3-strategy $q$-deformed evolutionary dynamics under the payoff matrix given in Eq.~(\ref{eq main: RPS payoff}) for $\delta=-2$ and $q=[0.2, 0.7, 1.5]$ in (a), (b) and (c) respectively. Green markers show stable fixed points.}
    \label{fig supp: delta_-2}
\end{figure}

For $q<q_{c}(-2)$ the centre point is a stable star. As shown in Sec.~\ref{appendix: corners} the corners in this region are sources. This leads to trajectories moving directly towards the centre point as illustrated in Fig.~\ref{fig supp: delta_-2}(a)

For $q_{c}(-2)\leq q< 1$ the centre point is now an unstable star. Again the corners in this region are sources. This leads to stable fixed points that are inside the simplex as illustrated by the green markers in Fig.~\ref{fig supp: delta_-2}(b). We note that stable internal fixed points could arise anywhere in the green region in Fig.~\ref{fig main: q delta phase diagram}, not just for $\delta=-2$. However, when $\delta$ moves further from $-2$ the centre becomes a stronger spiral sink/source, which forces trajectories into stable orbits rather than to stable fixed points.

For $q\geq 1$ the centre point is still an unstable star but the corners becomes sinks. This leads to trajectories moving directly away from the centre point, converging at the corners as illustrated in Fig.~\ref{fig supp: delta_-2}(c).

\FloatBarrier
\section{Fixation probability for \texorpdfstring{$q$}{}-deformed dynamics without replacement for \texorpdfstring{$2\times 2$}{} games} \label{appendix: fixation probability without replacement proof}
As in Sec.~\ref{appendix: fixation prob} we want to calculate the fixation probability $\phi_{i}$, except now for an adapted version of the model where we select an integer number $q$ of neighbours without replacement.

\begin{figure}[htbp]
    \centering
    \captionsetup{justification=justified}
    \includegraphics[scale=0.7]{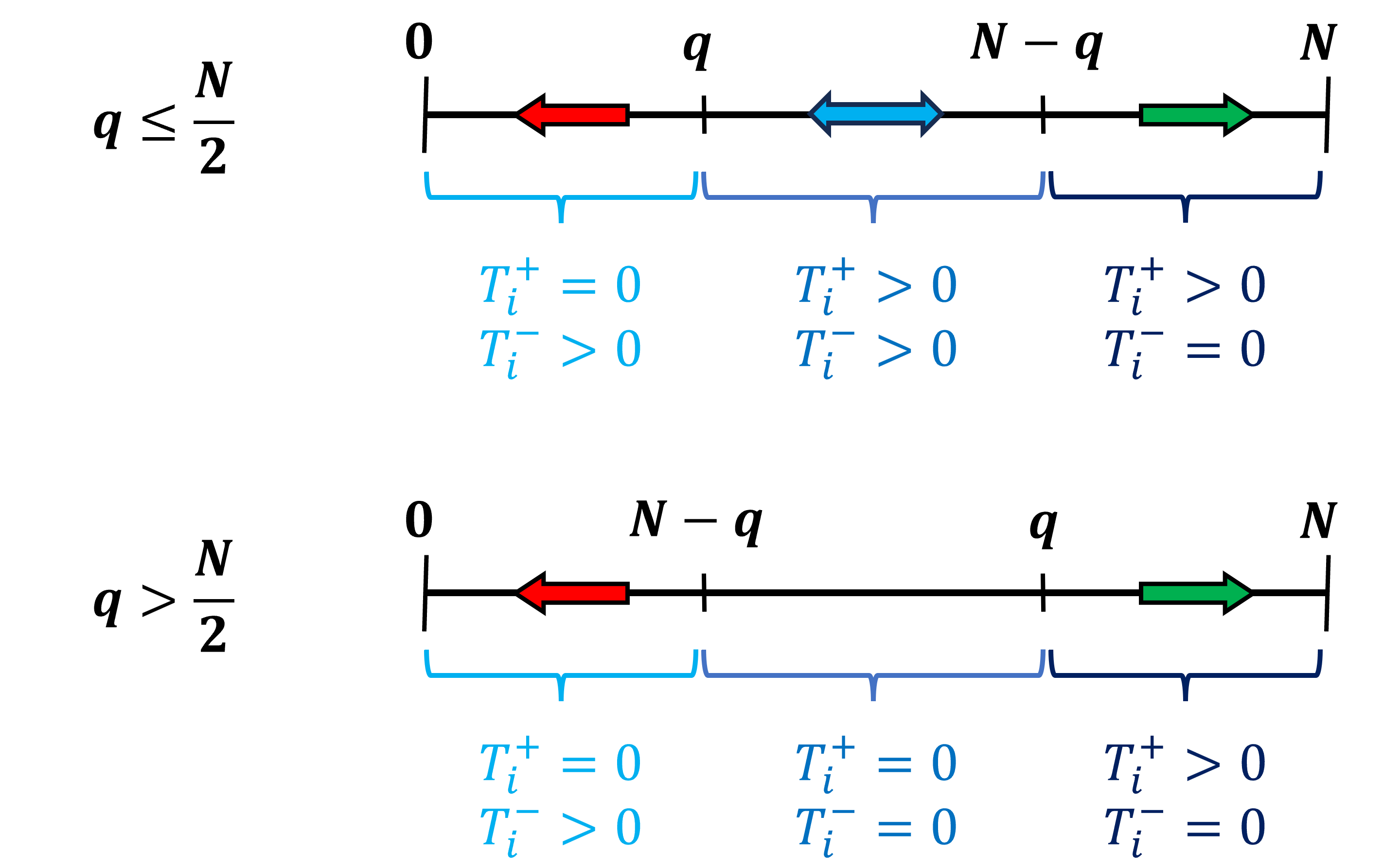}
    \caption{Demonstration of how the transition rates from Eqs. (\ref{eq supp: T plus finite}) and (\ref{eq supp: T minus finite}) change based on the number of type $A$ agents, $i$. Top panel shows the case $q\leq\frac{N}{2}$. For $i<q$ or $i>N-q$ the system is always driven to $i=0$ or $i=N$ respectively. For $q\leq i \leq N-q$ both $T^{+}_{i}$ and $T^{-}_{i}$ are finite so the system can move in either direction. The bottom panel shows the case for $q>\frac{N}{2}$. If $i\leq N-q$ or $i\geq q$ the system is always driven to $i=0$ or $i=N$ respectively. For $q< i < N-q$ both $T^{+}_{i}$ and $T^{-}_{i}$ are $0$ so the system does not move.}
    \label{fig: piecewise rates}
    \end{figure}

The rates for the model without replacement are defined as follows:
\begin{subequations}
\begin{align}
    T_{i}^{+} &= 
    \begin{cases}
    0, &0 \leq i < q \\
    (N-i)g^{+}_{i}\prod_{k=1}^{q}\left(\frac{i-k+1}{N-k}\right),  &q \leq i \leq N \\
    \end{cases} \label{eq supp: T plus finite} \\
    T_{i}^{-} &= 
    \begin{cases}
    ig^{-}_{i}\prod_{k=1}^{q}\left(\frac{N-i-k+1}{N-k}\right), &0 \leq i\leq N-q\\
    0, &N-q < i \leq N
    \end{cases} \label{eq supp: T minus finite}
\end{align}
\end{subequations}
Fig.~\ref{fig: piecewise rates} illustrates the rates for the cases of $q\leq\frac{N}{2}$ and $q>\frac{N}{2}$. We see that it is often the case that the system is driven to one of the absorbing states at $i=0$ or $i=N$. For $q\leq\frac{N}{2}$ and $q \leq i \leq N-q$ both $T^{+}_{i}$ and $T^{-}_{i}$ are non-zero so the system can move in either direction. For $q>\frac{N}{2}$ and $N-q < i < q$ both $T^{+}_{i}$ and $T^{-}_{i}$ are 0 so the system does not move. This is because we require at least $q$ $A$ or $B$ agents in order for a node to choose $q$ neighbours of the opposite type.
    
For the case $q\leq\frac{N}{2}$, the probability to fixate with all agents being $A$ type is trivially $0$ if $i<q$ and $1$ if $i>N-q$. For $q \leq i \leq N-q$ we write the recursive expression
\begin{equation}
    \phi_{i} = T_{i}^{-}\phi_{i-1}+T_{i}^{+}\phi_{i+1}+(1-T_{i}^{-}-T_{i}^{+})\phi_{i}.
\end{equation}
Rearranging this equation, and defining $y_{i}=\phi_{i}-\phi_{i-1}$, we find $y_{i+1} = \gamma_{i}y_{i}$, where $\gamma_{i}=T_{i}^{-}/T_{i}^{+}$ is the ratio of the transmission rates. Unlike in the model with replacement [Eq.~(\ref{eq: gamma})], $\gamma_{i}$ has a piecewise definition:
\begin{equation}
    \gamma_{i} = \begin{cases}
    \infty, &0 \leq i<q, \\
    \frac{i}{N-1}\frac{g^{-}(i)}{g^{+}(i)}\prod_{k=1}^{q}\left(\frac{N-i-k+1}{i-k+1}\right), &q\leq i \leq N-q \\
    0, &N-q < i \leq N
    \end{cases}. \label{eq: gamma no replacement}
\end{equation}
Noting that $\phi_{q}$ is the smallest non-zero $\phi$, we can write
\begin{align}
    y_{q} &= \phi_{q}-\phi_{q-1} = \phi_{q}, \nonumber \\
    y_{q+1} &= \gamma_{q}y_{q} = \gamma_{q}\phi_{q}, \nonumber \\
    &\vdots \nonumber \\
    y_{k} &= \phi_{q}\prod_{j=q}^{k-1}\gamma_{j}, \label{eq supp: no replacement yk defn}
\end{align}
for $k\geq q$. Now, assuming $i \geq q$ we can manipulate $\phi_{i}$ into the following form,
\begin{align}
    \phi_{i} &= (\phi_{q}-\phi_{q-1})+(\phi_{q+1}-\phi_{q})+...+(\phi_{i}-\phi_{i-1}) \nonumber \\
    &= y_{q} + y_{q+1} + ... + y_{i} \nonumber \\
    &= \sum_{k={q}}^{i}y_{k} \nonumber \\
    &= \phi_{q}\sum_{k=q}^{i}\prod_{j=q}^{k-1}\gamma_{j}, \label{eq: phi_i no replacement}
\end{align}
noting the first line is possible as all values other than $\phi_{q-1}$, which is zero anyway, cancel. In going from the third to fourth line we use Eq.~(\ref{eq supp: no replacement yk defn}). To determine $\phi_{q}$ we use the largest $\phi_{i}$ that we know to be equal to one, i.e. $\phi_{N-q+1}$,
\begin{gather}
    \phi_{N-q+1}=1=\sum_{k=q}^{N-q+1}y_{k}=\phi_{q}\sum_{k=q}^{N-q+1}\prod_{j=q}^{k-1}\gamma_{j}, \nonumber \\
    \implies \phi_{q} = \frac{1}{\sum_{k=q}^{N-q+1}\prod_{j=q}^{k-1}\gamma_{j}}. \label{eq: phi_q}
\end{gather}
Substituting Eq.~(\ref{eq: phi_q}) into Eq.~(\ref{eq: phi_i no replacement}) we get an expression for the fixation probability in the intermediate regime,
\begin{equation}
    \phi_{i} = \frac{\sum_{k=q}^{i}\prod_{j=q}^{k-1}\gamma_{j}}{\sum_{k=q}^{N-q+1}\prod_{j=q}^{k-1}\gamma_{j}}.
\end{equation}

When $q>\frac{N}{2}$ all cases are trivial. $i\leq N-q$ and $i\geq q$ give $\phi_{i}$ as $0$ or $1$ respectively as the system is forced to the absorbing states. The intermediate case of $N-q < i < q$ gives $\phi_{i}=0$, as the system cannot move. Overall then for $q>\frac{N}{2}$ we have $\phi_{i}=0$ for $0\leq i < q$, and $\phi_{i}=1$ for $q \leq i \leq N$. 

Combining all results we have for the fixation probability:
\begin{equation}
    \phi_{i} = 
    \begin{cases}
        0, &(q\leq \frac{N}{2} \; \mbox{and} \; 0 \leq i < q) \; \mbox{or} \; (q>\frac{N}{2} \; \mbox{and} \; 0\leq i < q)\\
        \frac{\sum_{k=q}^{i}\prod_{j=q}^{k-1}\gamma_{j}}{\sum_{k=q}^{N-q+1}\prod_{j=q}^{k-1}\gamma_{j}}, &q \leq \frac{N}{2} \; \mbox{and} \; q\leq i \leq N-q \\
        1, &(q\leq \frac{N}{2} \; \mbox{and} \; N-q <i\leq N) \; \mbox{or} \; (q>\frac{N}{2} \; \mbox{and} \; q\leq i\leq N)
    \end{cases}
\end{equation}

\end{document}